\documentclass[10pt,twocolumn,letterpaper]{article}

\usepackage[pagenumbers]{cvpr} 

\usepackage{multirow}
\usepackage{makecell}

\definecolor{cvprblue}{rgb}{0.21,0.49,0.74}
\usepackage[pagebackref,breaklinks,colorlinks,allcolors=cvprblue]{hyperref}

\def\paperID{*****} 
\def\confName{CVPR}
\def\confYear{2025}

\title{Confident Splatting: Confidence-Based Compression of 3D Gaussian Splatting via Learnable Beta Distributions}

\author{AmirHossein Naghi Razlighi\\
Sharif University of Technology\\
{\tt\small amirhossein.razlighi@sharif.edu}
\and
Elaheh Badali Golezani\\
Sharif University of Technology\\
{\tt\small elahe.badali@sharif.edu}
\and
Shohreh Kasaei\\
Sharif University of Technology\\
{\tt\small kasaei@sharif.edu}
}

\begin{document}
\twocolumn[{
    \maketitle
    \vspace{-2.5em}  
    \begin{center}
        \includegraphics[width=\linewidth]{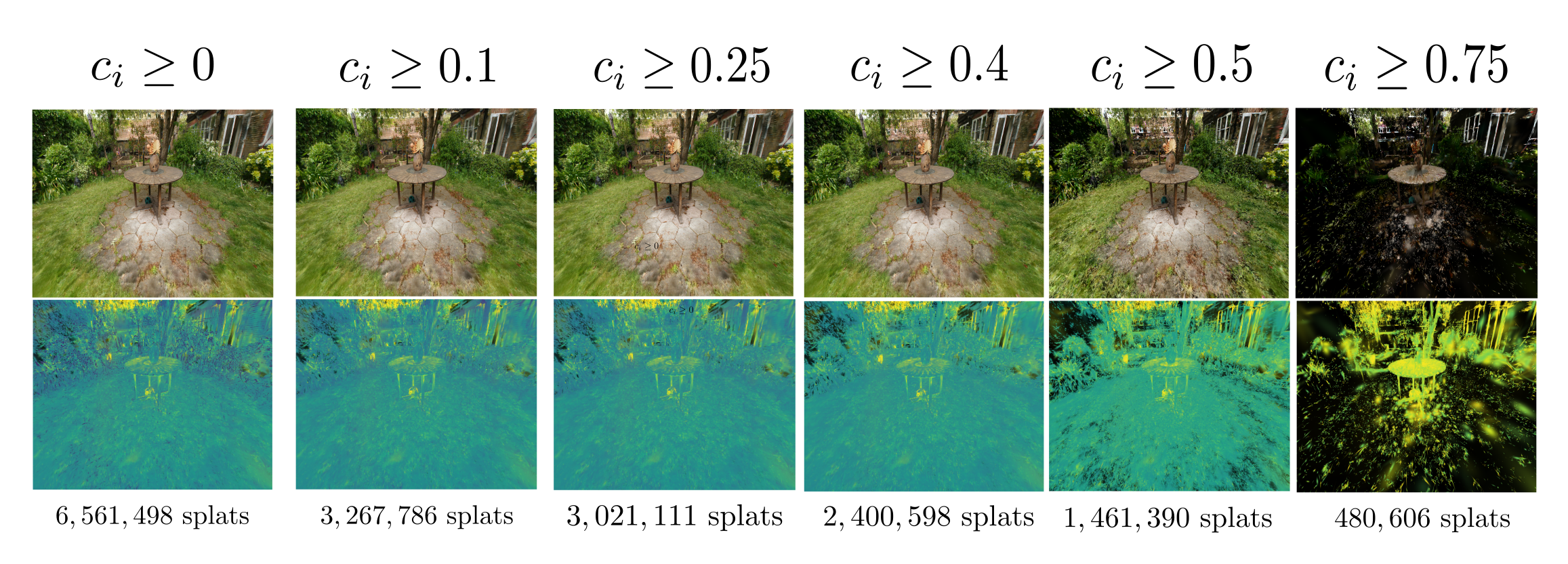}
        \captionof{figure}{Overview of proposded confidence-based Gaussian Splatting compression result, on the mip-nerf-360 garden \cite{mip-nerf-360} scene. Confidence scores are learned as Beta distributions and used for pruning.}
        \label{fig:overall_view}
    \end{center}
}]

\begin{abstract}
3D Gaussian Splatting enables high-quality real-time rendering, but often produces millions of splats, resulting in excessive storage and computational overhead. We propose a novel lossy compression method based on learnable confidence scores modeled as Beta distributions. Each splat's confidence is optimized through reconstruction-aware losses, enabling pruning of low-confidence splats while preserving visual fidelity. The proposed approach is architecture-agnostic and can be applied to any Gaussian Splatting variant. In addition, the average confidence values serve as a new metric to assess the quality of the scene. Extensive experiments demonstrate favorable trade-offs between compression and fidelity compared to prior work. Our code and data are publicly available \href{https://github.com/amirhossein-razlighi/Confident-Splatting}{here}.
\end{abstract}    
\section{Introduction}
\label{sec:intro}

3D Gaussian Splatting (3DGS)~\cite{gaussian-splatting} has revolutionized 3D scene representation and reconstruction due to its efficiency and quality. However, most scenes (especially those with medium or high complexity) are represented by millions of splats, which impose significant storage and rendering overhead. Before the advent of 3DGS, novel view synthesis (NVS) was primarily performed using Neural Radiance Field (NeRF) \cite{nerf}, which employs volumetric rendering \cite{volumetric-rendering} for differentiable rendering. This method produces high-quality results with strong theoretical foundations. However, its rendering pipeline is computationally intensive, making it unsuitable for real-time applications. Although some works like InstantNGP \cite{instant-ngp} tried to use HashGrid encoding with a much smaller MLP, it still incorporated the volumetric rendering, and thus had those limitations. In contrast, 3DGS adopts a point-based method that leverages GPU pipelines, which have been optimized over decades for rasterization. By using this form of differentiable rasterization, it achieves comparable visual quality with real-time rendering speeds.

One major drawback of 3DGS is its unbounded number of splats. By using adaptive filtering, the number of splats can easily reach millions for mediocre and complex scenes~\cite{liu2025maskgaussian}. Studies have shown that many of these splats are not crucial for the final visual fidelity. Some of these splats are designed to reconstruct very fine-grained details, which contribute virtually nothing to perceivable visual quality from the viewpoint of humans. A fraction of these splats also consists of floaters or artifacts that are not part of the original scene geometry~\cite{ali2025compression}.

These issues have opened up a research direction for compressing 3DGS scenes. Two branches have been made to achieve such goal. One is compressing splats' properties (such as Spherical Harmonics (SH) coefficients) without any loss on the final reconstruction. \cite{girish2024eagles, lee2025compression, ali2025compression} These approaches focus on representation mostly. The other method to  tackle this issue, is to consider some loss on the final reconstruction quality, as long as the quality-size tradeoff is significant enough. Our method lies in the second category. While similar methods in this category tried to tackle this problem, our method proposes a novel way for compressing (and also evaluating) the 3DGS scenes. We introduce a new method for estimating a confidence score for each splat. Our method is easily transferrable from one method of 3DGS to another (as we will demonstrate further) without any removal in the original pipeline. Via estimating parameters of a \textit{Beta distribution} over each splat, we will estimate its confidence score. Furthermore, we will optimize this property of splats by introducing useful loss functions and regularizations in the pipeline. Finally, after training is completed, our method can act as an evaluator (for how high-quality are the splats) and also act as a knob-based compression technique, giving the user the freedom to choose its optimal quality-size tuple for its own usage. In summary, we propose the following:

\begin{enumerate}
    \item A new confidence score estimation method is proposed which is based on estimating $\alpha$ and $\beta$ parameters and estimating a \textbf{Beta} distribution placed on each splat.

    \item  The sparsity loss, entropy loss, and ranking-based saliency loss are applied to optimize the $\alpha$ and $\beta$ parameters that determine the confidence scores. These confidence scores are also used in the rasterization process to further optimize the parameters’ values by integrating them into the opacity computation.
    \item After the training phase (which can be transferred to any 3DGS pipeline with minimum effort), the user can use the knob-based compression method by setting a threshold for the confidence of splats. This allows generating a compressed scene (by removing low-confidence splats) with minimal degradation in visual fidelity.
    \item It is shown that the proposed method can also serve as a metric to evaluate 3DGS methods across different scenes to understand and compare the quality of splats, where a higher average confidence indicates a higher-quality scene.
\end{enumerate}

\section{Related Works}
\label{sec:related_works}
In this section, we provide an overview of novel view synthesis and 3DGS, followed by a review of recent advancements in the field with a focus on compression.

\subsection{Novel view synthesis}
3D reconstruction is the process of creating a three-dimensional model from 2D images. In the 3D reconstruction field, Novel View Synthesis (NVS) \cite{xia2023survey} aims to generate images of a scene from viewpoints that were not part of the original observations, allowing virtual exploration from new perspectives. High resolution and sufficient multi-view input images are necessary for realistic novel view synthesis in order to guarantee precise visual reconstruction. Various machine learning techniques used for this purpose include epipolar geometry and stereo vision \cite{lee2024generalizable}, depth estimation \cite{cao2022fwd}, generative models \cite{chan2023generative}, and neural scene representations such as Neural Radiance Field (NeRF) \cite{nerf}. Nevertheless, each of these methods also has its drawbacks. Epipolar geometry and stereo vision are threatened by imprecise calibration and occlusions or textureless regions. Depth estimation fails to generalize well on different domains and complex scenes. Generative models produce a blurry output and are problematic for training and, NeRFs require dense views, lengthy training, and are costly computationally. Considering the restrictions imposed by the present methods in speed, scalability, and visual quality, there is an increase in the need for solutions that can deliver real-time performance along with high-quality synthesis.

\subsection{3D Gaussian Splatting}
3DGS \cite{gaussian-splatting} is a rendering and scene representation technique that models 3D environments using spatially distributed anisotropic Gaussian functions, referred to as Gaussian splats. Each splat holds the information about its location, geometry, and orientation (via covariance), color, and transparency, thus enabling an efficient and adaptive representation of the scene. To render the splats, a forward-splatting process is used, which enables them to blend seamlessly thus giving the result of photorealistic and very intricate reconstructions of the original scene.

3DGS is an explicit real-time method that makes it highly time-efficient during both training and testing. In addition, due to the continuous nature of Gaussian functions, smooth level-of-detail transitions and intrinsic anti-aliasing can be achieved, resulting in high visual quality across resolution changes and viewpoint variations \cite{chen2024survey}. These properties of 3DGS make it an appropriate approach for novel view synthesis, interactive scene exploration, and real-time rendering of large-scale environments.

3DGS generally requires a large number of Gaussians to model a scene, thus consuming a lot of memory and having computational overhead at training and rendering time. However, not all of these Gaussians contribute equally to the final reconstructed scene. In practice, only a fraction of the relevant Gaussians is ultimately required for good scene representation. As a result, there is significant potential for compression by identifying and retaining only the most relevant Gaussians, thereby improving efficiency without sacrificing visual fidelity \cite{chen2024survey}.

\subsection{Compression and Pruning in 3DGS}
Different methods have been proposed to prune and compress a 3DGS scene. These methods can be categorized into two main categories:
\begin{enumerate}
    \item \textbf{Changing the representation}: Several methods aim to modify the main representation of Gaussians or specific property of them such as spherical harmonics coefficients. Although these methods change the representations, these are lossless meaning that they preserve original visual quality. Various approachs have been proposed, such as using tri-planes, bilateral grids, and quantized embeddings \cite{lee2025compression, girish2024eagles, ali2025compression}. While these methods can achieve strong results compared to original 3DGS, they require changing the whole pipeline and representation, making them less scalable. Moreover, applying them to different methods and strategies of 3DGS requires significant effort.
    \item \textbf{Pruning the scene}: Different methods have been proposed to prune the 3DGS scene and reduce the number of splats per scene. These approaches can be applied either during training or at test time and may be either lossy or lossless. For instance, RadSplat \cite{radsplat} is a method for combining NeRF supervision on the point-based optimizations of 3DGS. During training, it defines an importance score based on ray contribution, and changes the Adaptive filtering process by adding an extra check in the training (removing splats with importance less than a predefined threshold). While this method achieves state-of-the-art performance compared to methods such as original 3DGS \cite{gaussian-splatting}, ZipNeRF \cite{zipnerf}, and Baked SDF \cite{baked-sdf}, it requires changing the pipeline and also it needs re-training for every new threshold, unlike our method, which can be seamlessly added on top of any existing pipeline. Mini-splatting \cite{mini-splatting} leverages similar approach for pruning. It also provides a new approach instead of using Adaptive filtering, which refines the distribution of Gaussian centers in the scene (which is out of scope of our study in this paper). Another method, LP3DGS \cite{lp3dgs} learns the optimal pruning ratio during training using a Gumbel-Sigmoid–based importance metric. This method is a train-time method and does not provide a user-controllable trade-off between visual quality and scene size. In contrast, our method not only outperforms all of these approaches but also operates as a test-time method. It offers users full flexibility to choose their desired balance between visual fidelity and scene compression.
\end{enumerate}
\section{Prepared Dataset}
\label{sec:data}

To evaluate proposed method across a diverse range of scenarios, we constructed and curated several scenes through two complementary approaches.
\paragraph{YouTube-Reconstructed Scenes.}
We gathered public videos from YouTube depicting well-known landmarks, such as the Eiffel Tower. From these videos, high-quality frames were extracted at regular intervals and Structure-from-Motion (SfM) was performed to reconstruct sparse 3D point clouds. These reconstructions were subsequently used as initialization for generating full 3DGS scenes. This setup allows us to validate proposed method under unconstrained, in-the-wild conditions with varying lighting, motion, and occlusion.
\paragraph{Confidence-Enhanced Scene Dataset.}
In addition to custom scenes, we applied our method on a diverse set of standard 3DGS benchmark datasets. For each processed scene, we released the full set of Gaussians, augmented with our optimized confidence scores. This curated dataset provides a rich resource for downstream tasks, including confidence-guided mesh reconstruction, uncertainty-aware view synthesis, and automatic pruning strategies. We hope it facilitates future research on compression, quality estimation, and the interpretability of point-based neural representations.

All datasets, preprocessing scripts, and trained scenes with confidence annotations will be publicly available upon acceptance. You can find more info on the dataset details in supplementary material \ref{sec:supp_dataset}.

\section{Proposed Method}
\begin{figure*}[th]
    \centering
    \includegraphics[width=\linewidth]{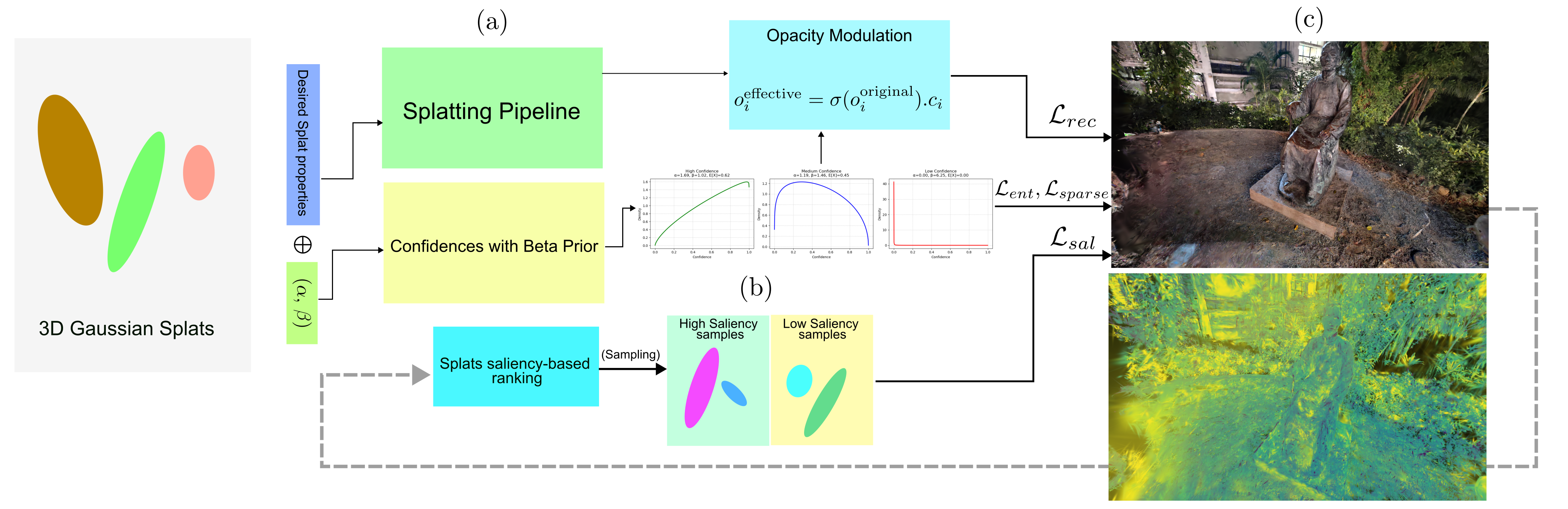}
    \caption{Overview of \textit{"confident splatting"} method. We extend any 3DGS pipeline by introducing learnable confidence parameters $(\alpha, \beta)$ for each splat, modeled by a Beta distribution. \textbf{(a)} The original pipeline remains unchanged, except for an added \textbf{opacity modulation} using confidence values. \textbf{(b)} Confidence parameters are optimized via entropy, sparsity, and saliency losses; sample distributions are shown. \textbf{(c)} Final renderings (and optional confidence heatmaps) are produced via standard rasterization. Saliency (image-space gradients) is stored and used for detached \textbf{saliency-based ranking} (Dashed arrows indicate non-differentiable paths).}
    \label{fig:architecture}
\end{figure*}

\label{sec:method}
We propose a method for learning per-splat confidence scores to enable both \textit{lossy compression} and \textit{scene quality assessment} in 3DGS. These scores are derived from a learnable \textit{Beta distribution} and are optimized jointly with the reconstruction loss. During training, the confidence affects rendering; post-training, it supports user-defined pruning for efficient storage and rendering. You can see different components of our pipeline, working alongside any desired 3DGS pipeline in Fig. \ref{fig:architecture}.

\subsection{Confidence Estimation with Beta Distributions}
For confidence score estimation of each splat, unlike previous methods, we do not regress a simple confidence($c_i$) value. Instead, a Beta distribution is assumed for each splat, represented by two parameters $\alpha$ and $\beta$. The Beta distribution parameterized by $\alpha$ and $\beta$ is defined as follows:

\begin{align}
        f_{X}(x ; \alpha, \beta) &= \frac{\Gamma(\alpha + \beta)}{\Gamma(\alpha).\Gamma(\beta)} x^{\alpha - 1}.(1-x)^{\beta - 1} \nonumber \\
    &= \frac{1}{B(\alpha,\beta)} . x^{\alpha - 1}. (1 - x)^{\beta - 1}
\end{align}

Furthermore, the expected value of $X \sim Beta(\alpha, \beta)$ can be calculated as follows:
\begin{equation}
\begin{aligned}
     E[X] &= \int_{-\infty}^{\infty} x.f_X(x) dx \\
     & = \int_{-\infty}^{\infty} \frac{1}{B(\alpha, \beta)} . x^{(\alpha + 1) - 1}. (1-x)^{\beta - 1} dx
     \\
     &= \frac{1}{B(\alpha, \beta)}. B(\alpha + 1, \beta) = \frac{\Gamma(\alpha + \beta)}{\Gamma(\alpha).\Gamma(\beta)} . \frac{\Gamma(\alpha + 1) . \Gamma(\beta)}{\Gamma(\alpha + \beta + 1)} \\
     &= \frac{\alpha}{\alpha + \beta}
\end{aligned}
\end{equation}

\begin{figure*}[t]
    \centering
    \includegraphics[width=\linewidth]{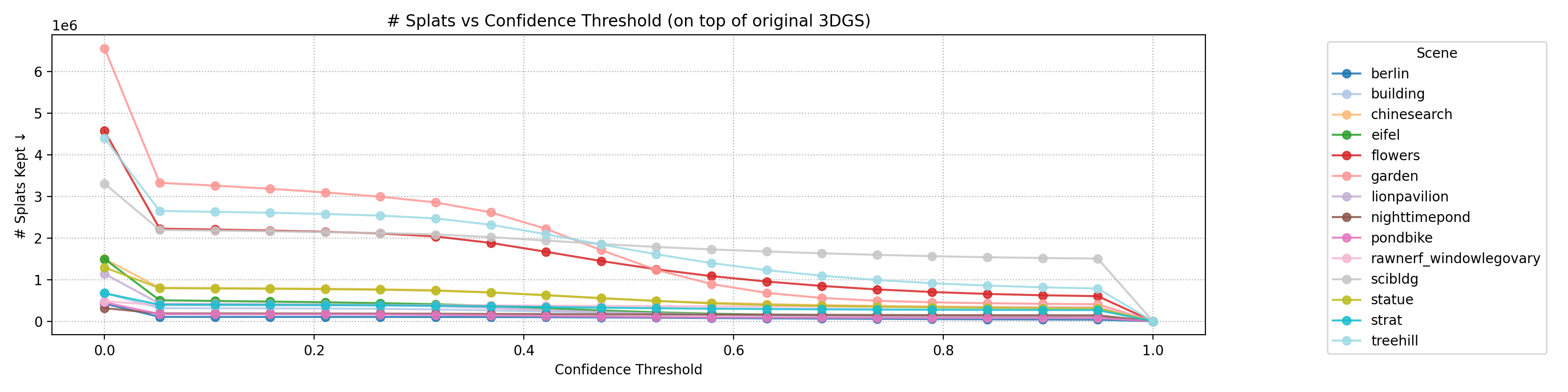}
    \caption{Compression evaluation on different scenes, to show how confidence thresholding affects number of kept splats in the scene (and thus, reduces the size of the whole scene, proportionally).}
    \label{fig:compression_vs_num_splats}
\end{figure*}

So, for each splat, a Beta distribution is fitted to its confidence value (parameterized by $\alpha$ and $\beta$ which are optimized within the pipeline), and the confidence is computed as follows:

\begin{equation}
    c_i = E[C_i] = \frac{\alpha_i}{\alpha_i + \beta_i} \quad ;C_i \sim Beta(\alpha_i, \beta_i)
\end{equation}

\subsection{Loss Functions and Optimization}
To optimize our confidence score parameters, we leverage using three regularizers, alongside whatever loss function the main pipeline is using. For instance, the original 3DGS \cite{gaussian-splatting} pipeline uses SSIM loss and RGB L1 loss, denoted as \textit{"Reconstruction loss"}. We have:

\begin{equation}
    \mathcal{L}_{total} = \mathcal{L}_{rec} + \lambda_1 . \mathcal{L}_{sparse} + \lambda_2 . \mathcal{L}_{ent} + \lambda_3 . \mathcal{L}_{sal}
\end{equation}

Where \textbf{sparsity loss} encourages fewer active splats. "Active Splats" denotes the splats with confidence more than $50\%$ ($c_i >= 50\%$)
\begin{equation}
    \mathcal{L}_{sparse} = \frac{1}{N} \sum_{i=1}^{N} c_i
\end{equation}
Also, \textbf{negative entropy loss} avoids overconfidence in splats. We know that the low-entropy for our beta distribution mean that our distribution is very peaked near $0$ or $1$, thus means that the $c_i$ is overconfident in the fact that that specific splat is useful ($c_i=1$) or useless($c_i=0$). On the other hand, high entropy for our beta distribution means the distribution is uncertain (flat, near uniform). So, we penalize low entropy to discourage the optimizer from pushing all $\alpha \gg \beta$ ($confidence \rightarrow 1$) or pushing $\beta \gg \alpha$ ($confidence \rightarrow 0$)\\

\begin{equation}
    \mathcal{L}_{\text{ent}} = \mathbb{E}_i[-\mathcal{H}(Beta(Softplus(\alpha_i), Softplus(\beta_i))]
\end{equation}

Finally, \textbf{Saliency ranking loss} is proposed. Most of the methods that are using confidence scores, use absolute values to guide confidence values through optimization (like image gradients). On the other hand, we use a \textit{relative} approach towards confidence calculation. Let's take the highest loss (in image space) splats (easily obtained via image gradients) and the lowest loss splats. We sample $\mathcal{P}$ of them (the higher the $P$, the more accurate is the optimization, but the training will be slower due to more computation). So, $\mathcal{P}$ is a set of saliency-ranked splat index pairs. Our ranking-based saliency loss ensures that splats with higher contribution to gradients are more confident in comparison with splats with lower contributions

\begin{equation}
    \mathcal{L}_{\text{sal}} = \frac{1}{K} \sum_{(i,j) \in \mathcal{P}} \max(0, 1 + c_j - c_i)
\end{equation}
Where $i$ is the index for splat with higher contribution to gradients and $j$ is index for splat with lower contribution.

\subsection{Rasterization with Confidence Modulation}
It is obvious that confidence scores of each splat should be somehow related to the final rendered image. From the concept of confidence, we expect that higher confidence splats contribute more in the rendered image (and be more impactful) than the ones with lower confidence. phenomena is modeled by re-calculating opacity before the rasterization step:
\begin{equation}
    o^{\text{effective}}_{i} = \sigma(o^{\text{original}}_i).c_i
\end{equation}
In this way, each splat’s opacity is modulated by its current confidence score before rasterization and thus, the confidence score is directly integrated into the rendering process.

\subsection{Post-Training Pruning and User-Controlled Compression}
Note that we are not explicitly removing any splat based on its confidence in the training process. Confidence values are estimated and optimized (via our beta distribution prior for each splat) and stored as properties for each splat. After training process has finished, a user can simply use these confidence values as a knob-based compression and remove splats less than a certain confidence score based on his needs. The user can choose the threshold which is the best for him in terms of \textit{"keeping high visual fidelity"} Vs \textit{"lowering the number of splats as possible"}.

\subsection{Transferability and Compatibility}
It should be mentioned that the proposed method do not rely on the base pipeline of 3DGS anywhere of our confidence estimation process. The only thing that matters is that the pipeline should have base properties of splats (including SH coefficients or RGB colors, Rotation, Scale, and Opacity). Also, this method does not change any part of the pipeline, it only adds a new property to each splat. The only minor affect it may have on the base pipeline is that the opacity calculation is enhance via confidence values. In the Experiments section \ref{sec:experiments} we will show that this change not only does not degrade the quality, but also improves quality metrics in some cases for different 3DGS methods.
\section{Experiments}
\label{sec:experiments}

We conduct extensive experiments to evaluate the effectiveness of proposed confidence-based compression scheme for 3DGS. Our goals are to assess:
(1) compression-quality trade-offs,
(2) generalizability across datasets, and
(3) compatibility with other 3DGS variants.

\subsection{Experimental Setup}

\begin{figure*}[t]
    \centering
    \begin{subfigure}[b]{0.48\linewidth}
        \centering
        \includegraphics[width=\linewidth]{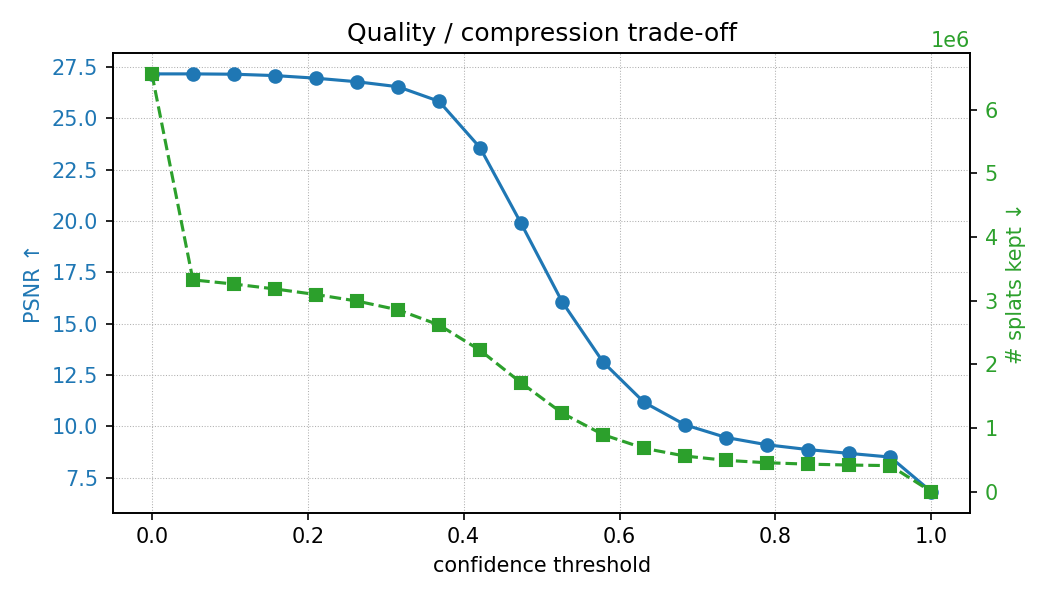}
        \caption{Quantitative plot: PSNR and number of splats vs threshold.}
        \label{fig:garden_quantitative}
    \end{subfigure}
    \hfill
    \begin{subfigure}[b]{0.48\linewidth}
        \centering
        \includegraphics[width=\linewidth]{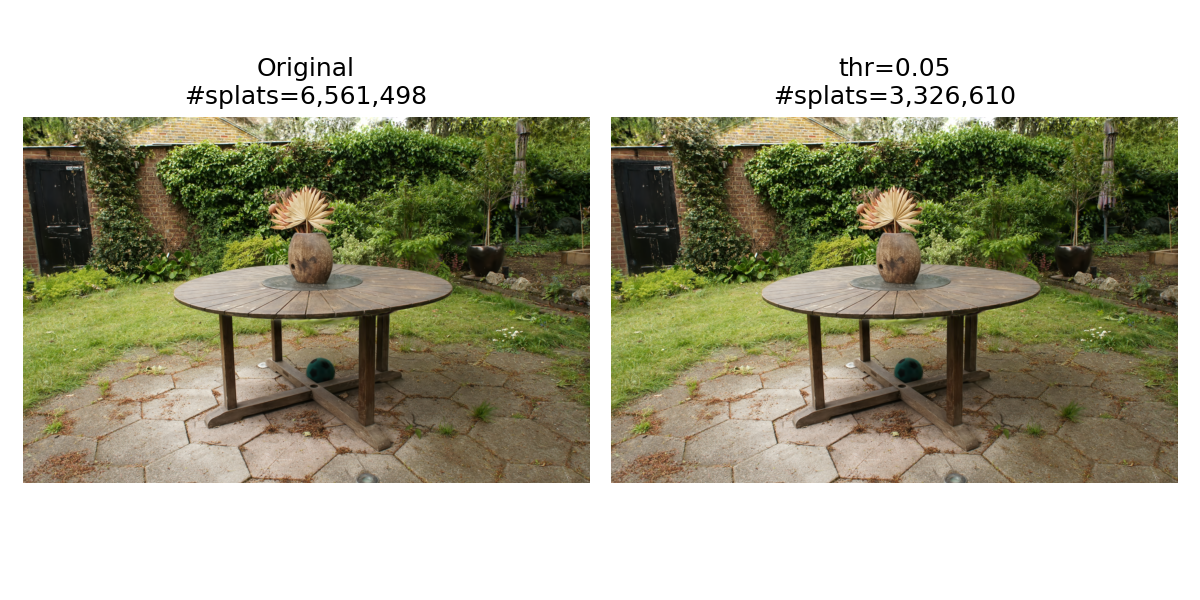}
        \caption{Visual comparison: original vs a sample confidence-pruned.}
        \label{fig:garden_quality_side_by_side}
    \end{subfigure}
    \caption{Compression evaluation on the Garden scene. (a) metric trends. (b) rendering with and without pruning.}
    \label{fig:garden_scene_base}
\end{figure*}

We evaluate our method on a diverse suite of scenes from various datasets:
\begin{itemize}
    \item \textbf{BILARF} test scenes: \textit{building}, \textit{chinesearch}, \textit{lionpavilion}, \textit{nighttimepond}, \textit{pondbike}, \textit{statue}, \textit{strat}
    \item \textbf{Custom Eiffel Tower Scene:} constructed from YouTube footage via frame extraction and SfM (see Section \ref{sec:data})
    \item \textbf{MipNeRF-360}: \textit{garden}
    \item \textbf{MipNeRF-360-extra}: \textit{flowers}, \textit{treehill}
    \item \textbf{ZipNeRF dataset}: \textit{berlin} scene
    \item \textbf{Tanks\&Temples dataset}: \textit{truck} and \textit{train} scenes.
\end{itemize}

\vspace{1em}
For each scene, we train both the \textbf{baseline 3DGS} model \cite{gaussian-splatting} and our proposed method on top of it, which augments splats with learnable Beta-distributed confidence scores. After training, splats are pruned below confidence threshold $\tau$, sweep across a range of $\tau$ values, and evaluate rendering quality and scene sparsity.

To demonstrate compatibility with other 3DGS pipelines, the same protocol is repeated using the \textbf{MCMC-GS} \cite{mcmc} variant. We report results for:
\begin{itemize}
    \item \textbf{MCMC-GS (baseline)}
    \item \textbf{Ours@MCMC-GS}: The proposed method applied on top of MCMC-GS, optimizing confidence scores alongside the base model
\end{itemize}

\subsection{Quantitative Results: Compression vs Quality}

the compression-quality trade-off curves are plotted by sweeping confidence thresholds and measuring PSNR on the test views. For each threshold, we record:
\begin{itemize}
    \item Number of retained splats
    \item PSNR, SSIM, and LPIPS scores
\end{itemize}

\begin{figure}[h]
    \centering
    \begin{subfigure}[b]{0.95\linewidth}
        \includegraphics[width=\linewidth]{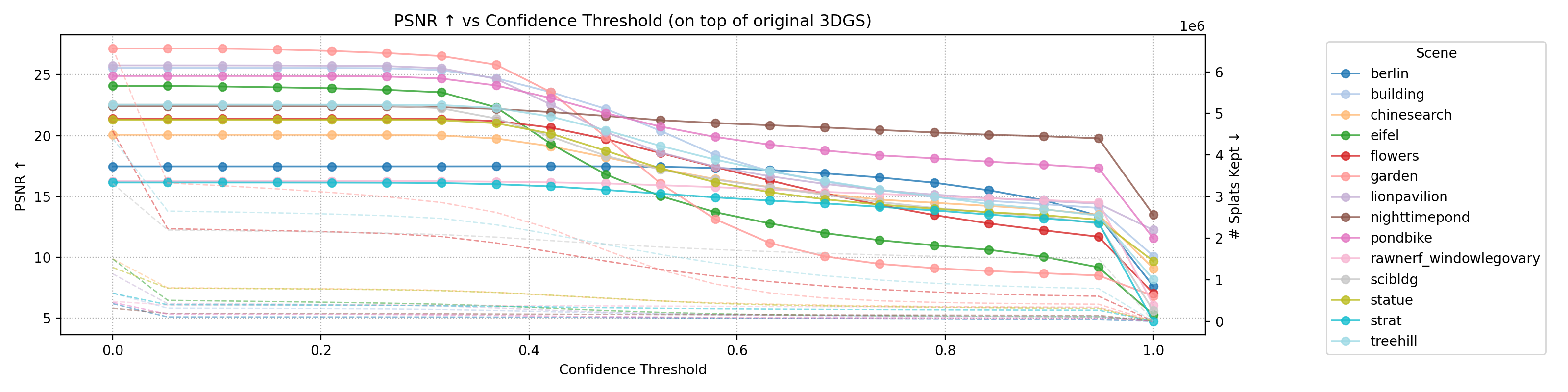}
        \caption{PSNR ↑}
        \label{fig:psnr_vs_thresh}
    \end{subfigure}
    \hfill
    \begin{subfigure}[b]{0.95\linewidth}
        \includegraphics[width=\linewidth]{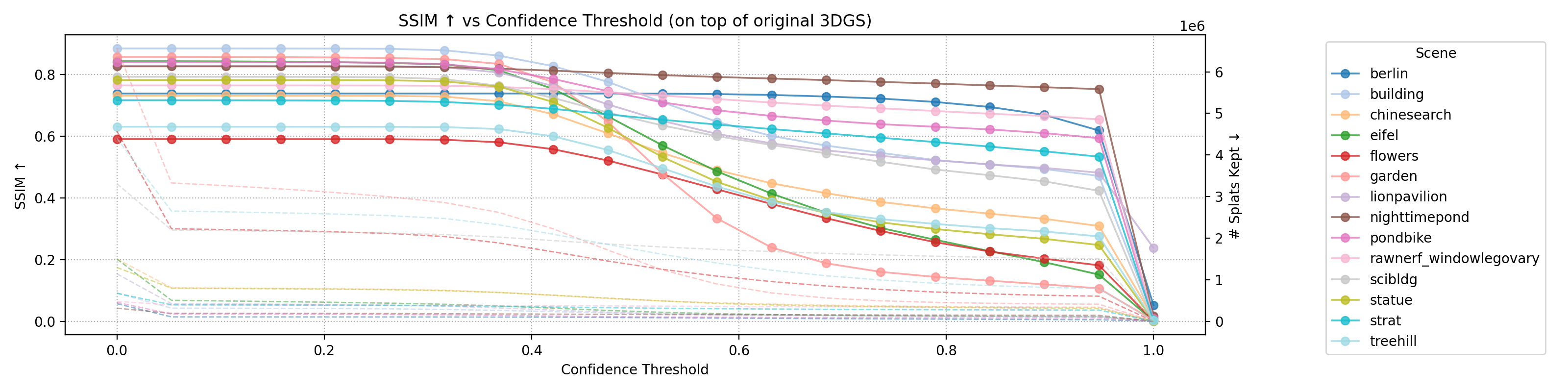}
        \caption{SSIM ↑}
        \label{fig:ssim_vs_thresh}
    \end{subfigure}
    \hfill
    \begin{subfigure}[b]{0.95\linewidth}
        \includegraphics[width=\linewidth]{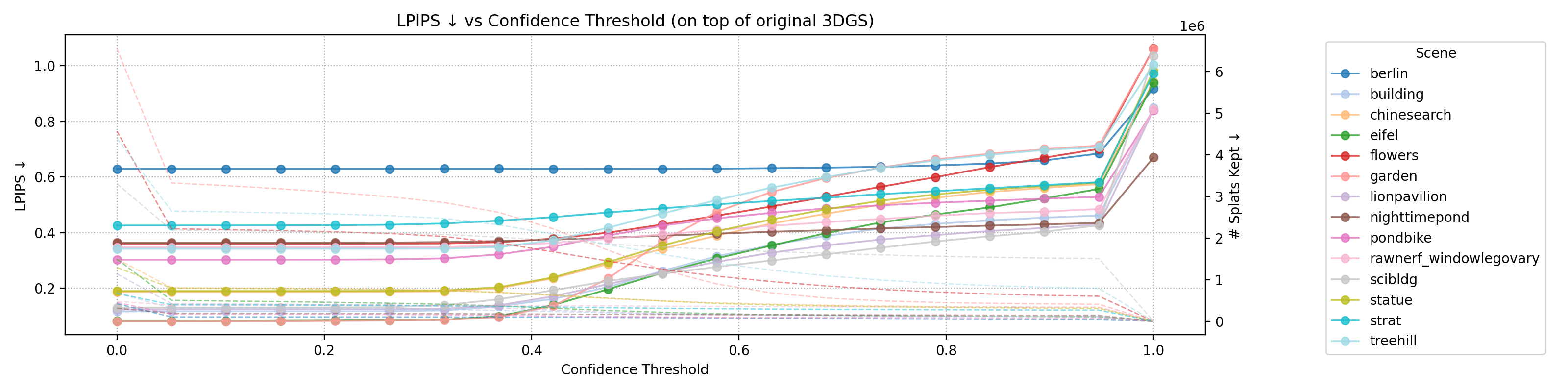}
        \caption{LPIPS ↓}
        \label{fig:lpips_vs_thresh}
    \end{subfigure}

    \caption{Effect of confidence-based pruning on visual quality metrics across different datasets. Note how metrics degrade slowly while number of splats drops drastically.}
    \label{fig:compression_vs_quality}
\end{figure}

As seen in Figure \ref{fig:compression_vs_quality}, our method achieves high PSNR with a substantially reduced number of splats. Compared to the original 3DGS, our confidence scores provide a better signal for identifying redundant or low-impact splats. In most scenes, number of splats drops to half (Fig. \ref{fig:compression_vs_num_splats}) while the quality metrics like PSNR do not change or change with a really small tolerance (Fig. \ref{fig:compression_vs_quality}). 

\subsection{Qualitative Results}

\begin{figure*}[h]
    \centering
    \includegraphics[width=\linewidth]{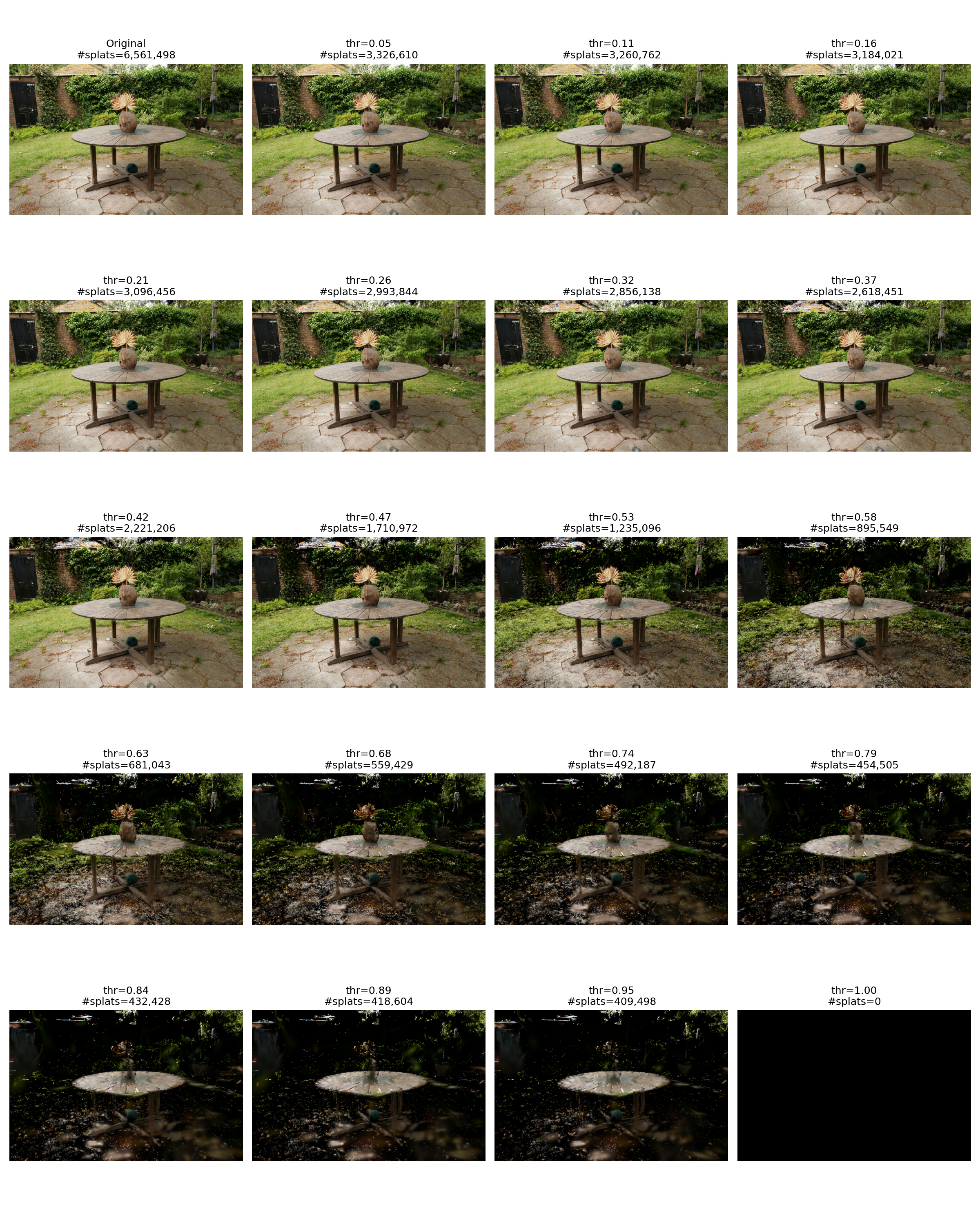}
    \caption{Qualitative renderings at varying confidence thresholds. Our confidence-aware pruning preserves visual fidelity while discarding low-impact splats.}
    \label{fig:garden_scene_qualititative}
\end{figure*}

To visually assess our method, we provide renderings at various compression thresholds. In Figure \ref{fig:garden_scene_qualititative}, you can see renderings of a sample scene (\textit{Garden scene from MipNeRF-360 dataset \cite{mipnerf}}), while we sweep and prune the splats with confidence values under a certain threshold. Number of kept splats is also written in these plots.

In Fig \ref{fig:garden_scene_base}, more than half of the splats are pruned in the garden scene, with less than $0.5$ decrease in PSNR metric. Also, this fact (near zero degradation in the first confidence pruning values) can be seen in the right plot, where we have plotted the original scene and a sample pruned scene (with confidence value $5\%$).

\subsection{Comparison}
\begin{table*}[tbh]
    \centering
    \resizebox{\linewidth}{!}{%
    \begin{tabular}{llccccc}
        \toprule
        \textbf{Scene} & \textbf{Method} & \textbf{PSNR ↑} & \textbf{SSIM ↑} & \textbf{LPIPS ↓} & \textbf{\#Gaussians ↓} & \textbf{SQR↓} \\
        \midrule
        \multirow{8}{*}{Flowers}
            & 3DGS        & $21.450$ & $0.596$ & $0.345$ & $3,590,000$ & $0.1433$ \\
            & Rad-Splat @ 50\%       & $21.367$ & $0.589$ & $0.361$ & $1,751,775$ & $0.0757$ \\
            & Mini-Splatting @ 50\%  & $21.382$ & $0.5871$ & $0.360$ & $3,384,239$ & $0.1366$ \\
            & Ours\texttt{@}Base3DGS @100\%    & $21.383$ & $0.5907$ & $0.360$ & $4,572,755$ & $0.1761$ \\
            & Ours\texttt{@}Base3DGS @$\sim$50\%     & $21.383$ & $0.5906$ & $0.360$ & $2,227,814$ & $0.0943$ \\
            & Ours\texttt{@}MCMC @100\%    & $21.592$ & $0.6050$ & $0.341$ & $1,000,000$ & $0.0442$ \\
            & Ours\texttt{@}MCMC @$\sim$95\%    & $21.334$ & $0.5941$ & $0.347$ & $963,860$ & $0.0432$ \\
            & \textbf{Ours\texttt{@}MCMC @$\sim$90\%}    & $20.614$ & $0.5684$ & $0.360$ & $915,488$ & \textbf{0.0425} \\
        \midrule
        
        \multirow{8}{*}{Eiffel Tower} 
            & 3DGS        & $22.877$ & $0.7969$ & $0.138$ & $600,000$ & $0.2077$ \\
            & Rad-Splat @ 50\%       & $22.524$ & $0.7919$ & $0.281$ & $395,873$ & $0.1484$ \\
            & Mini-Splatting @ 50\%  & $22.189$ & $0.7884$ & $0.283$ & $803,662$ & $0.2658$ \\
            & Ours\texttt{@}Base3DGS @100\%    & $24.077$ & $0.843$ & $0.082$ & $1,496,522$ & $0.3833$ \\
            & \textbf{Ours\texttt{@}Base3DGS @$\sim$27\%}     & $23.552$ & $0.8329$ & $0.087$ & $408,564$ & \textbf{0.1478} \\
            & Ours\texttt{@}MCMC @100\%    & $24.653$ & $0.8585$ & $0.072$ & $1,000,000$ & $0.2885$ \\
            & Ours\texttt{@}MCMC @$\sim$95\%    & $24.653$ & $0.8585$ & $0.072$ & $950,000$ & $0.2781$ \\
            & Ours\texttt{@}MCMC @$\sim$90\%    & $24.593$ & $0.8581$ & $0.072$ & $904,098$ & $0.2688$ \\
        \bottomrule
    \end{tabular}
    }
    \vspace{0.5em}
    \caption{A compact view of our comparisons among different methods and pruning ratios. @$X\%$ means that specific method, is keeping $X\%$ of the splats initially in its scene. For a more comprehensive comparison, please refer to the supplementary materials Tab. \ref{tab:comparison_complete}}
    \label{tab:comparison_compact}
\end{table*}

Alongside basic metrics for assessing quality (PSNR, SSIM, LPIPS) and size (number of splats in the scene), we introduce a new metric to show how well is the quality-compression trade-off. This metric is between $0$ and $1$ and the lower it is, the better and should be worse if the quality (represented by PSNR) is low ($0$ at its lowest). It should be minimized if for a certain PSNR, number of splats is decreased. The proposed formulation is as follows:

\begin{equation}
    \text{\textbf{SQR}} = \frac{num\_of\_splats}{num\_of\_splats + PSNR \times scale}
\end{equation}
Here, \textbf{SQR} stands for \textbf{\textit{Splats-to-Quality Ratio}}. Also, $scale$ is the order of number of splats in the original scene (without any compression). It can vary from $1e3$ to $1e6$. For example, if a base 3DGS scene (without any compression or pruning) has $3,000,000$ splats, the scale would be $1,000,000$. This is to normalize the importance of \textit{quality} and \textit{number of splats} at the same time, without one being largely more important than the other.

\subsection{Transferability Across Pipelines}
To validate the generalization of our method, we apply it to the MCMC-GS \cite{mcmc} variant. As shown in Table \ref{tab:mcmc_results}, the method achieves similar compression-quality tradeoffs when applied to MCMC-GS, confirming its compatibility with other splatting-based rendering pipelines. The point of our pipeline is (as it can be seen in Fig \ref{fig:architecture}) that it is fully transferable and makes no changes to the underlying logic of the method (like specific adaptive filtering strategies, rendering methods, etc.). It can be interpreted as appending new properties to splats which are being calculated in their own specific way. After training, the degree of compression (pruning) of unimportant splats can be controlled based on the calculated confidence scores for each splat in the scene.
\begin{table}[hbt]
    \centering
    \resizebox{\linewidth}{!}{%
    \begin{tabular}{llccc}
        \toprule
        \textbf{Scene} & \textbf{Method} & \textbf{PSNR@Orig} & \textbf{PSNR@95\%} & \textbf{PSNR@90\%} \\
        \midrule
        \multirow{2}{*}{Building} 
            & MCMC-GS & 26.09 & -- & -- \\
            & \textbf{Ours@MCMC-GS} & \textbf{25.96} & \textbf{25.62} & \textbf{24.39} \\
        \midrule
        \multirow{2}{*}{Flowers} 
            & MCMC-GS & 21.65 & -- & -- \\
            & \textbf{Ours@MCMC-GS} & \textbf{21.59} & \textbf{21.27} & \textbf{20.19} \\
        \midrule
        \multirow{2}{*}{Eiffel Tower} 
            & MCMC-GS & 24.57 & -- & -- \\
            & \textbf{Ours@MCMC-GS} & \textbf{24.65} & \textbf{24.65} & \textbf{24.59} \\
        \bottomrule
    \end{tabular}
    }
    \vspace{0.5em}
    \caption{Compression results on MCMC-GS \cite{mcmc} scenes. We report PSNR at original resolution, and after pruning $5\%$ and $10\%$ of total splats. All scenes originally contain $1,000,000$ splats.}
    \label{tab:average_confidence_results}
\end{table}

\subsection{Scene Quality Assessment Metric}

Finally, we investigate how the average confidence score of a scene correlates with its quality. It can be observed that scenes with fewer floaters/artifacts tend to exhibit higher mean confidence. Thus, our confidence scores may serve as a metric for evaluating reconstruction quality or even guiding automated scene refinement. The quality assessment metric can be defined as follows:

\begin{equation}
    ACS = \frac{1}{N} \sum_{i=1}^N c_i
\end{equation}

\begin{table*}[t]
    \centering
    \resizebox{\linewidth}{!}{%
    \begin{tabular}{llcccc}
        \toprule
        \textbf{Scene} & \textbf{Method} & \textbf{PSNR@Orig} & \textbf{SSIM@Orig\%} & \textbf{LPIPS@Orig\%} & \textbf{ACS ↑} \\
        \midrule
        \multirow{2}{*}{Building} 
            & Ours@MCMC-GS & 25.96 & 0.9057 & 0.0863 & \textbf{0.9391} \\
            & Ours@3DGS & 25.55  & 0.8844 & 0.1169 & 0.3063 \\
        \midrule
        \multirow{2}{*}{Flowers} 
            & Ours@MCMC-GS & 21.59 & 0.6049 & 0.3413 & \textbf{0.9236} \\
            & Ours@3DGS & 21.38 & 0.5906 & 0.3604 & 0.3087 \\
        \midrule
        \multirow{2}{*}{Eiffel Tower} 
            & Ours@MCMC-GS & 24.65 & 0.8585 & 0.7221 & \textbf{0.6712} \\
            & Ours@3DGS & 24.08 & 0.8431 & 0.0819 & 0.1874 \\
        \midrule
        \multirow{2}{*}{Truck} 
            & Ours@MCMC-GS & 25.82 & 0.8840 & 0.0888 & \textbf{0.8105} \\
            & Ours@3DGS & 25.21 & 0.8725 & 0.1034 & 0.2383 \\
        \midrule
        \multirow{2}{*}{Train} 
            & Ours@MCMC-GS & 22.61 & 0.8285 & 0.1445 & \textbf{0.7846} \\
            & Ours@3DGS & 21.80 & 0.8017 & 0.1764 & 0.3434 \\
        \midrule
        \multirow{2}{*}{Treehill} 
            & Ours@MCMC-GS & 22.96 & 0.6456 & 0.3520 & \textbf{0.9184} \\
            & Ours@3DGS & 22.53 & 0.6308 & 0.3421 & 0.3967 \\
        \midrule
        \multirow{2}{*}{Garden}
            & Ours@MCMC-GS & 26.80 & 0.8390 & 0.1176 & \textbf{0.8548} \\
            & Ours@3DGS & 27.16 & 0.8572 & 0.0814 & 0.2628 \\
        \midrule
        \bottomrule
    \end{tabular}
    }
    \vspace{0.5em}
    \caption{Comparing MCMC method \cite{mcmc} (with less floater and artifacts) and original 3DGS method \cite{gaussian-splatting} (with more artifacts and lower quality) in terms of scene quality (ACS).}
    \label{tab:mcmc_results}
\end{table*}
Here, \textbf{ACS} stands for \textit{"Average Confidence Score"} and $c_i$ stands for the confidence score of $i$th splat. By comparing the scenes in our experiments, both in quantitative and qualitative manners, we can see that for the same scene, the method with higher quality (PSNR) and lower artifacts and floaters, has higher \textit{ACS} in comparison with other scenes (with higher number of floaters and/or lower quality in terms of PSNR or SSIM or LPIPS).

\section{Conclusion and Future direction}
An efficient method to estimate confidence scores accurately using beta distributions fitted on top of each splat in the scene is proposed in this paper. Our method gives the freedom to users to choose their required compression-quality trade-off in \textbf{test-time} with no additional re-training or extra computations needed. These scenes (trained with different methods) with estimated confidence scores can be used for other research directions and problems too. It is worth noting that our confidence estimation framework addresses two of the most challenging problems in 3DGS: main object extraction and floater removal. Please refer to supplementary material for more explanations on these matters (\ref{sec:main_obj_extraction}, \ref{sec:folaters_removal}).
\clearpage
{
    \small
    \bibliographystyle{ieeenat_fullname}
    \bibliography{main}
}

\clearpage
\setcounter{page}{1}
\maketitlesupplementary

\section{YouTube Gathered Dataset}
\label{sec:supp_dataset}
Public videos from YouTube are gathered depicting well-known landmarks, such as the Eiffel Tower. From these videos, high-quality frames are extracted at regular intervals and Structure-from-Motion (SFM) is performed to reconstruct sparse 3D point clouds. By using these SFM point-clouds and also the main convention of dataset creation (with different image resolutions and also camera distortions) in the 3DGS research area, we created scenes with high diversity and also real-world conditions such as people walking around, high details in weather, trees and environment, and also camera not moving in a simple cyclic path. You can access more details on our dataset and how to access it publicly \href{here}{}. Also, the channels and videos that are used from YouTube are cited and linked in our repository. We have used \textit{"Walking Tour"} videos from YouTube channels like \href{https://www.youtube.com/watch?v=x7u70e9iQKQ&pp=ygUZZWlmZmVsIHRvd2VyIHdhbGtpbmcgdG91cg%3D%3D}{LADmob} and \href{https://www.youtube.com/watch?v=-30GJVgY1c8&pp=ygUibG91dnJlIG11c2V1bSBvdXRzaWRlIHdhbGtpbmcgdG91cg%3D%3D}{The CamFam Adventures (1)}, \href{https://www.youtube.com/watch?v=-30GJVgY1c8}{The CamFam Adventures (2)}, and \href{https://www.youtube.com/watch?v=-YE5aiF_gis}{Alta Walk}. \textit{"Confident Splatting Dataset"} contains \textbf{four large scale, in-the-wild scenes}:
\begin{enumerate}
    \item \textbf{Perspolis Scene}: Consists of $114$ images of a part of Takht-e-jamshid monument in Iran. It does not contain any serious distractors, but learning its surfaces and texture is hard since they are almost identical walls covered with soil and some historical carvings. 
    \item \textbf{Eiffel Tower Scene}: Consists of $1108$ images from Efiel Tower. This scene is one of the hardest scenes to reconstruct among in-the-wild scenes, since it has many distractors (people moving around in front of of the Eiffel Tower). Also, it is a large scale scene (containing trees and the environment in front of and behind of the main structure) and serious changes in camera movement.
    \item \textbf{Big Ben Scene}: Consists of $213$ images from the Big Ben landmark in London. This scene also contains a lot of distractors like people, telephone booth, and buses moving around. It starts from a distant view of the Big Ben, and goes toward it in a non-straight path.
    \item \textbf{Louvre Museum Scene}: Consists of $95$ images from the outside view of the building of Louvre Museum. This scene contains people moving around (as distractors) in front of the main structure. Also, the glass texture of the main structure alongside light reflections and refractions make this a hard scene for reconstructing its color and structure correctly. Also, the task of "main object prediction" in this scene is hard since this scene consists of many building around the main structure.
\end{enumerate}

\begin{figure}[tb] \centering
    \includegraphics[width=0.108\textwidth]{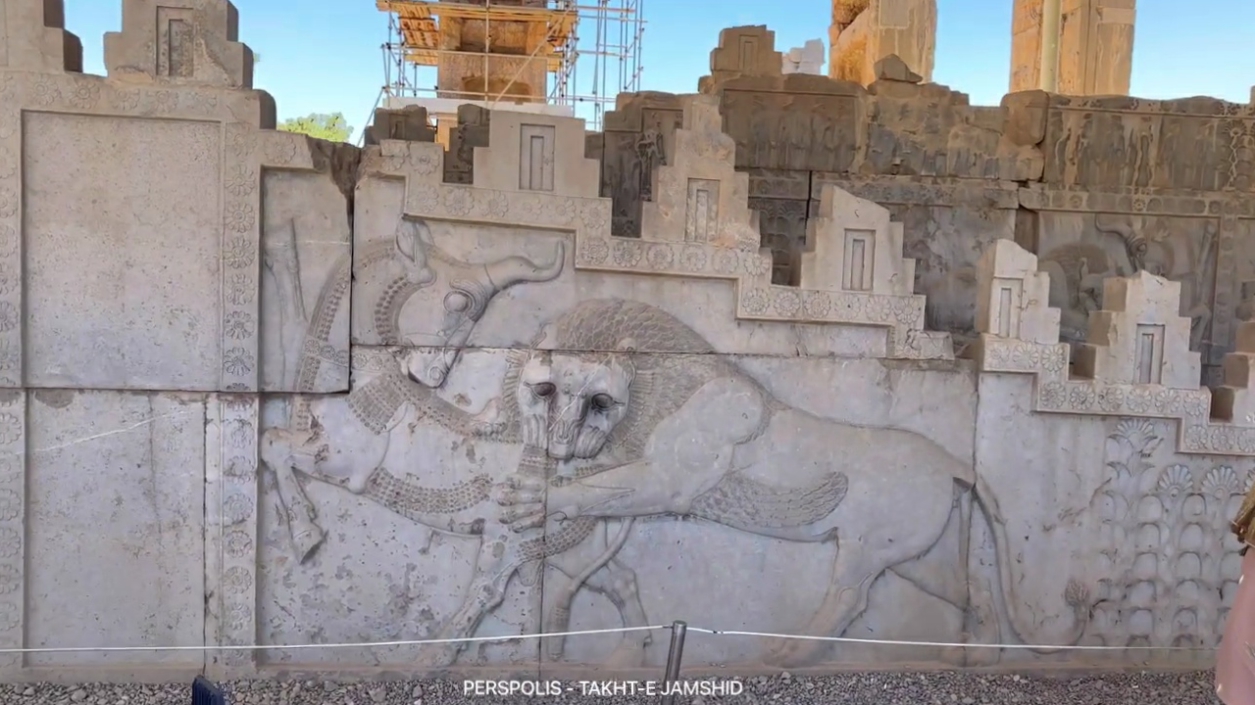}
    \includegraphics[width=0.108\textwidth]{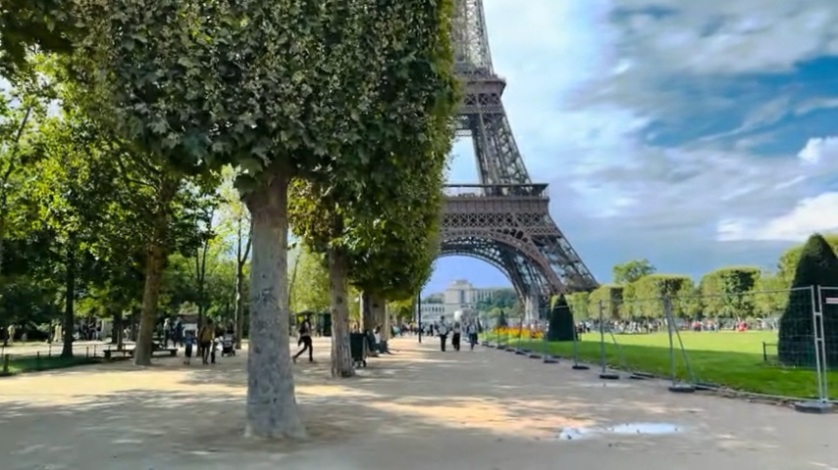}
    \includegraphics[width=0.108\textwidth]{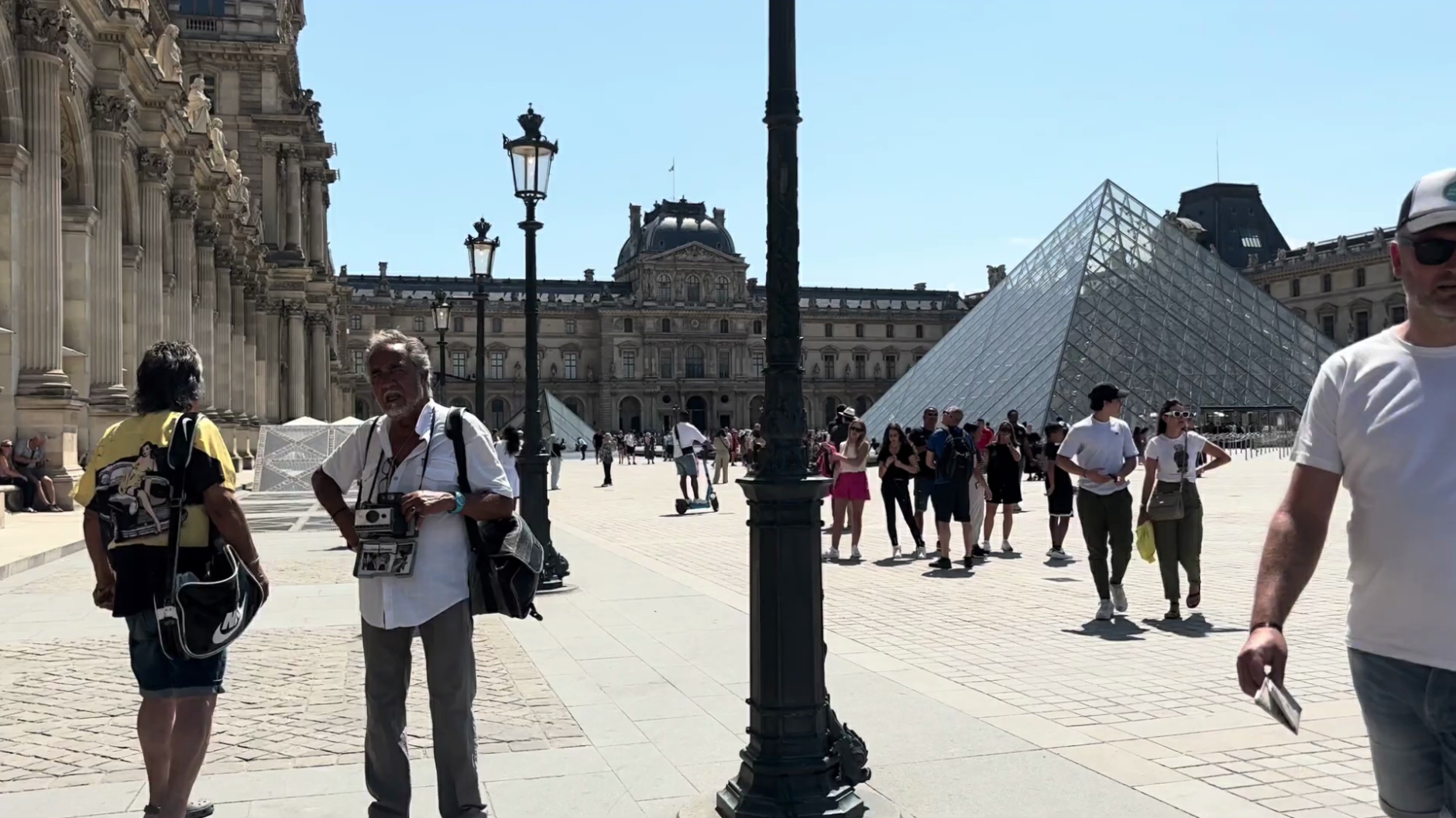}
    \includegraphics[width=0.108\textwidth]{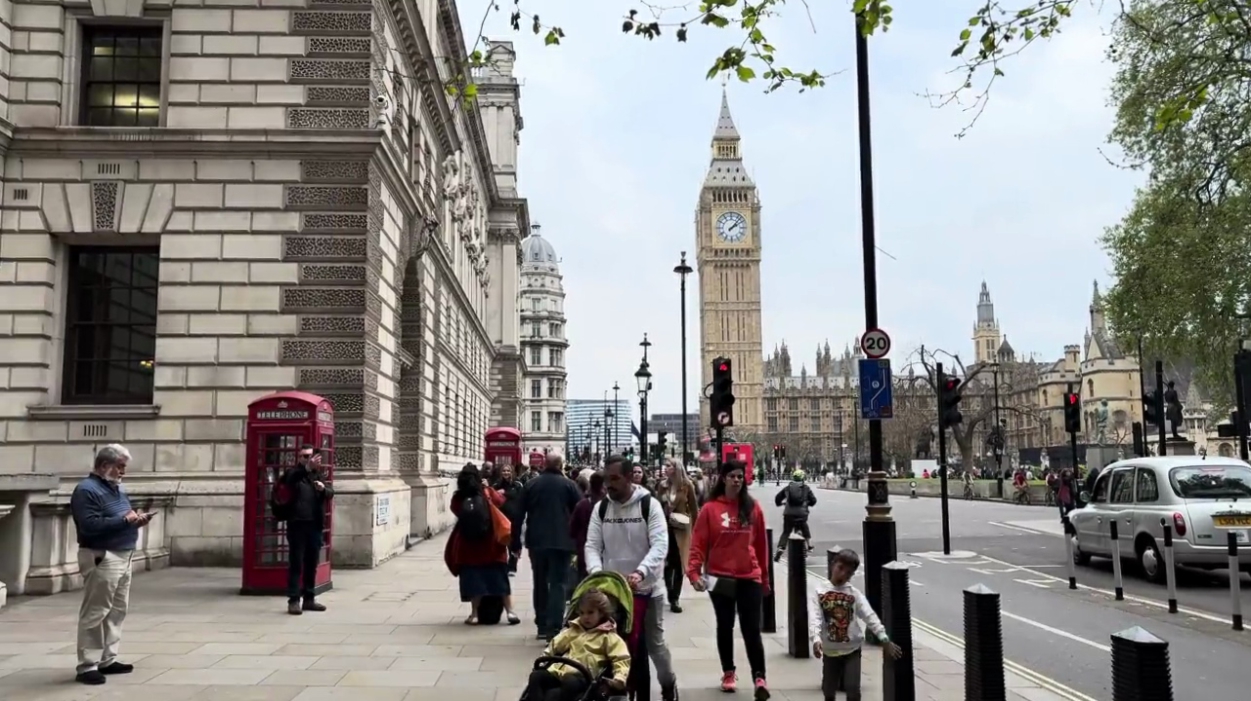}
    \\
    \includegraphics[width=0.108\textwidth]{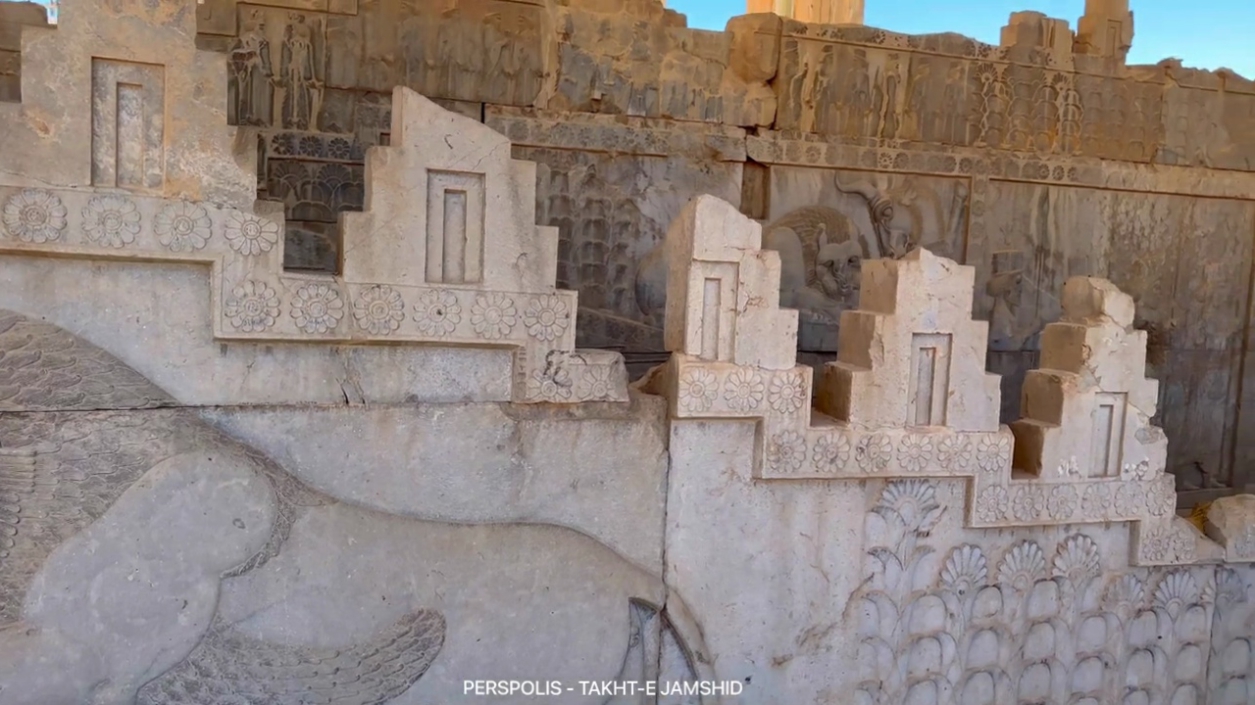}
    \includegraphics[width=0.108\textwidth]{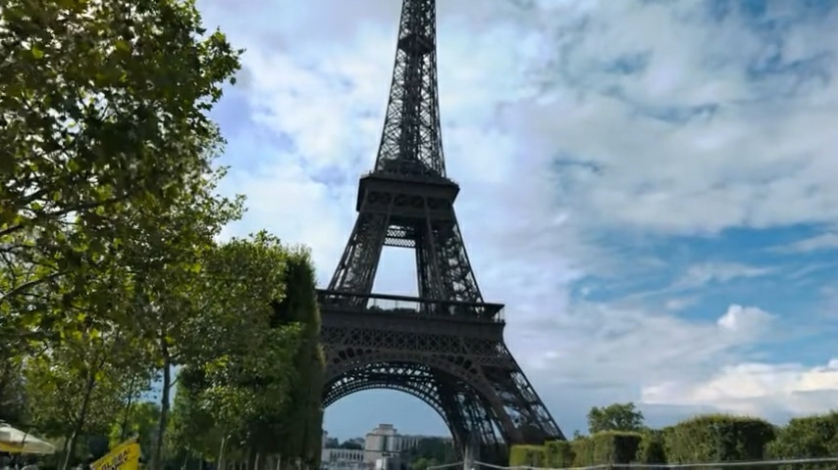}
    \includegraphics[width=0.108\textwidth]{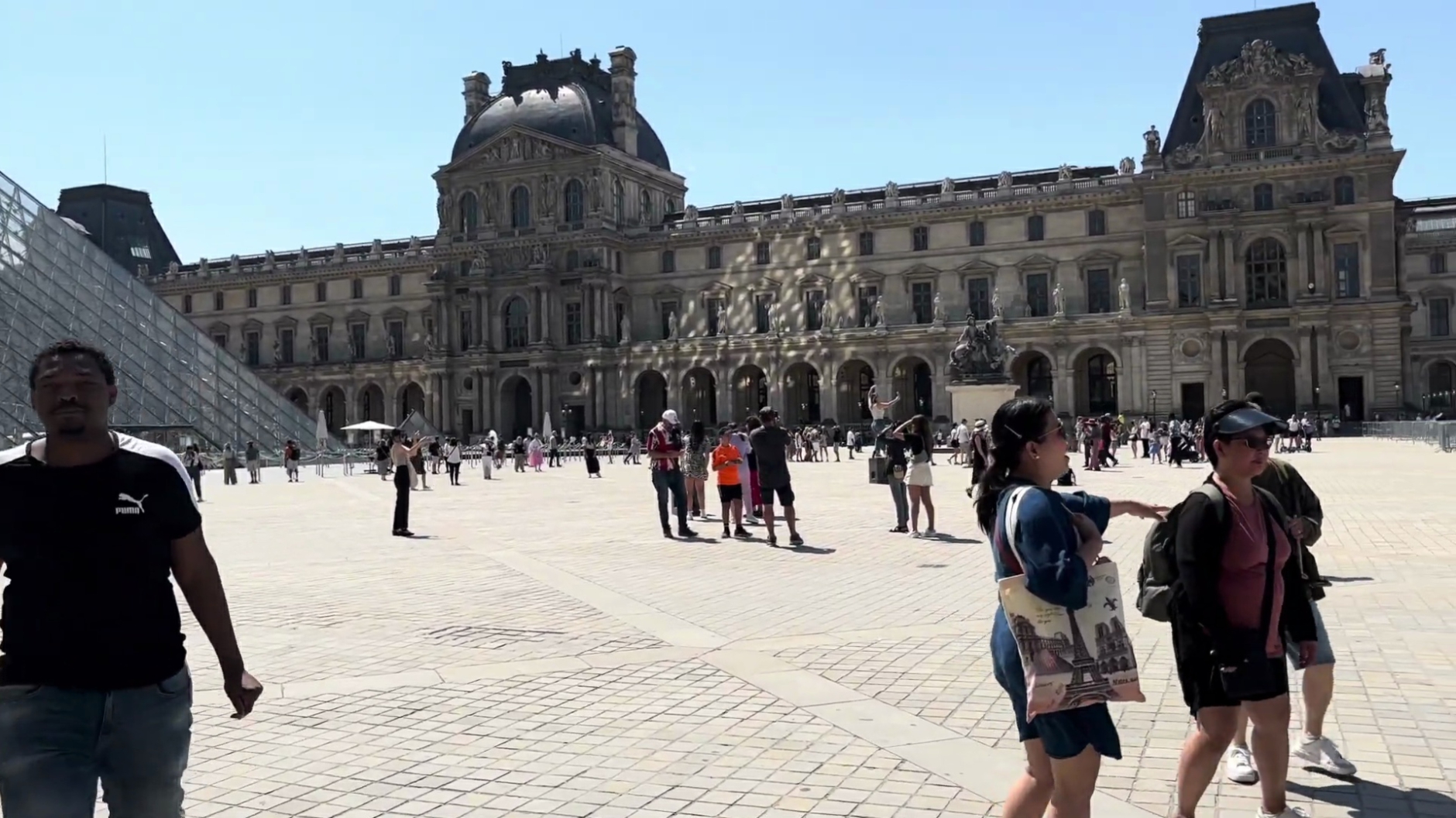}
    \includegraphics[width=0.108\textwidth]{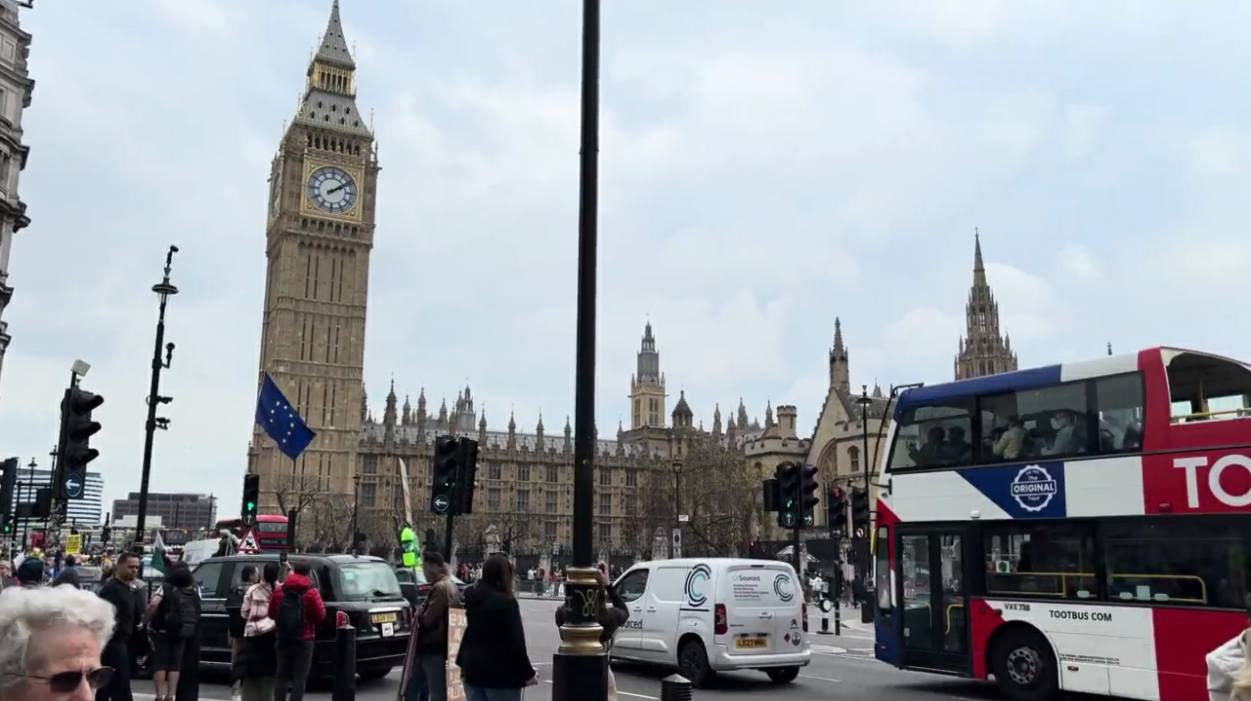}
    \\
    \includegraphics[width=0.108\textwidth]{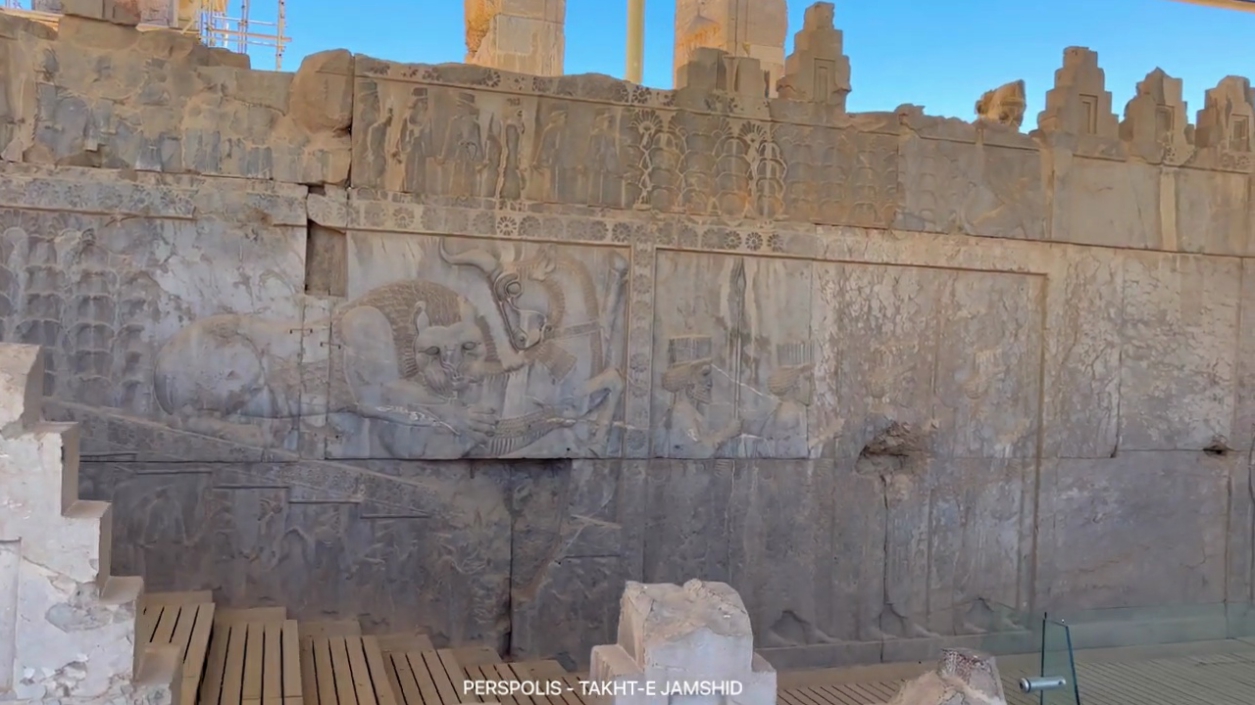}
    \includegraphics[width=0.108\textwidth]{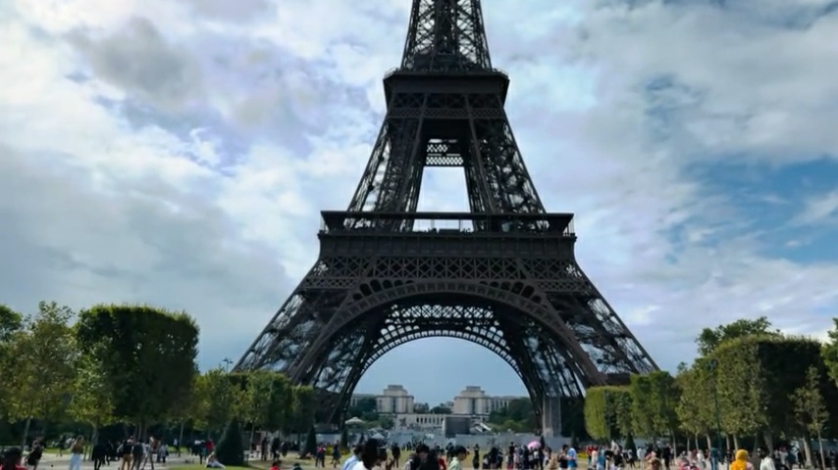}
    \includegraphics[width=0.108\textwidth]{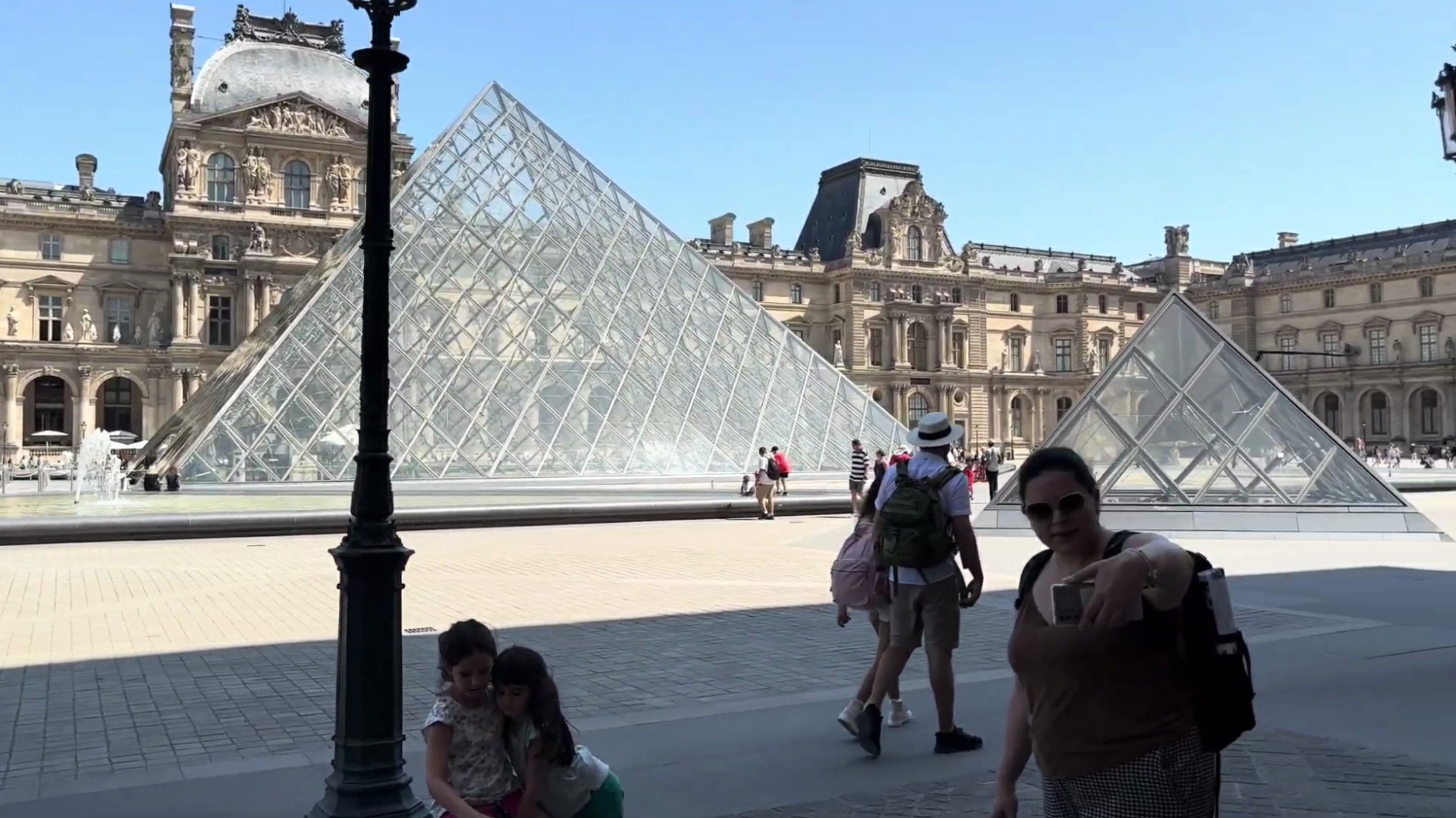}
    \includegraphics[width=0.108\textwidth]{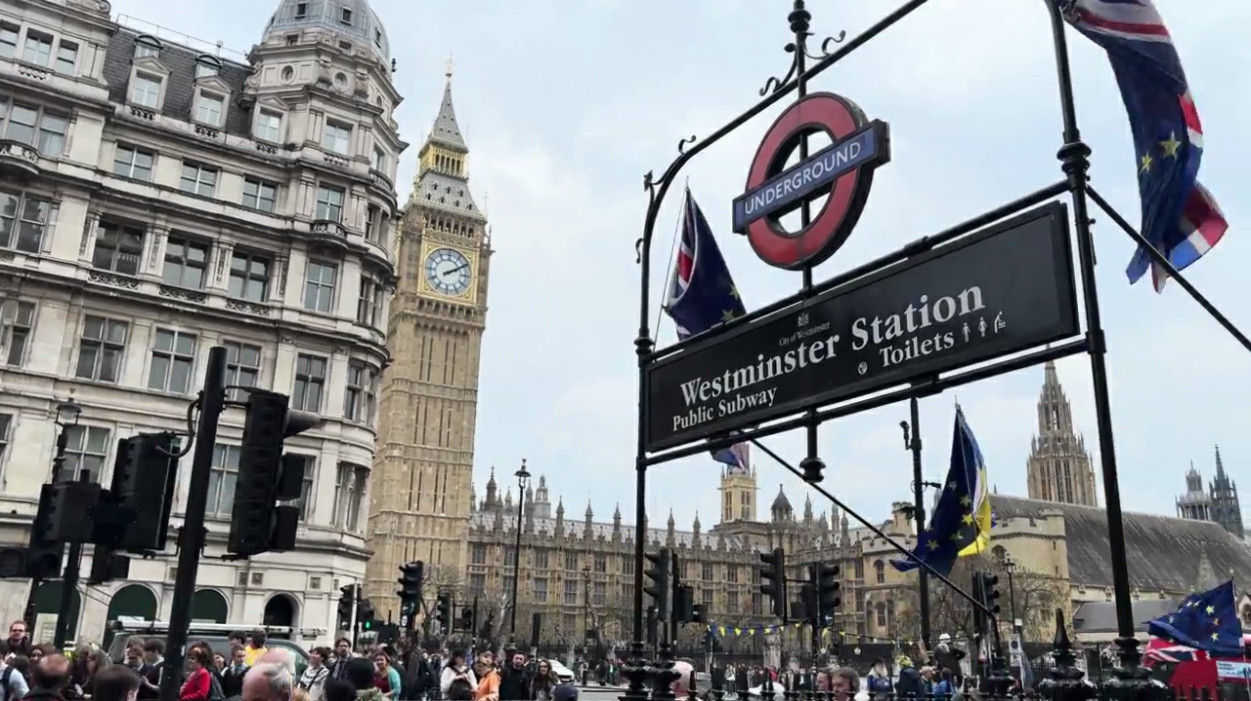}
    \\
    \includegraphics[width=0.108\textwidth]{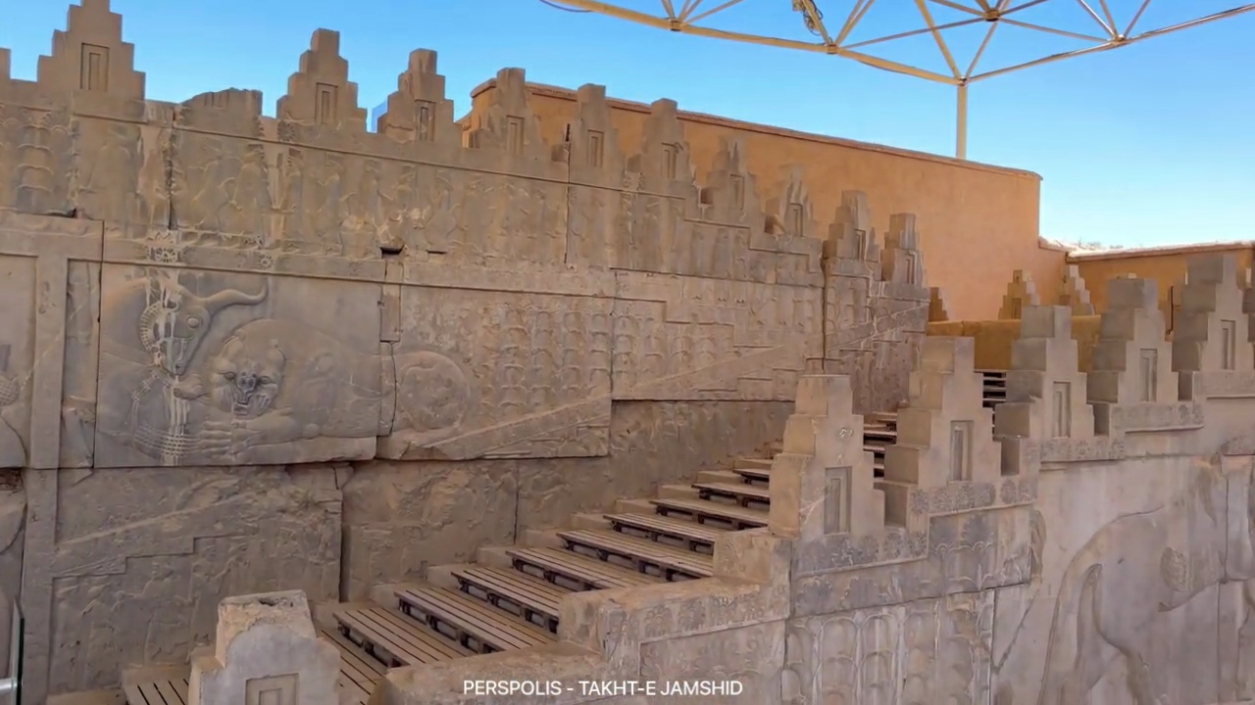}
    \includegraphics[width=0.108\textwidth]{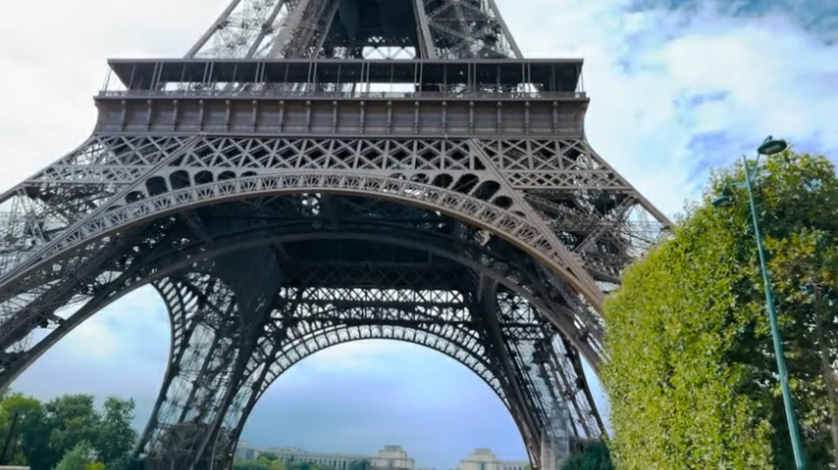}
    \includegraphics[width=0.108\textwidth]{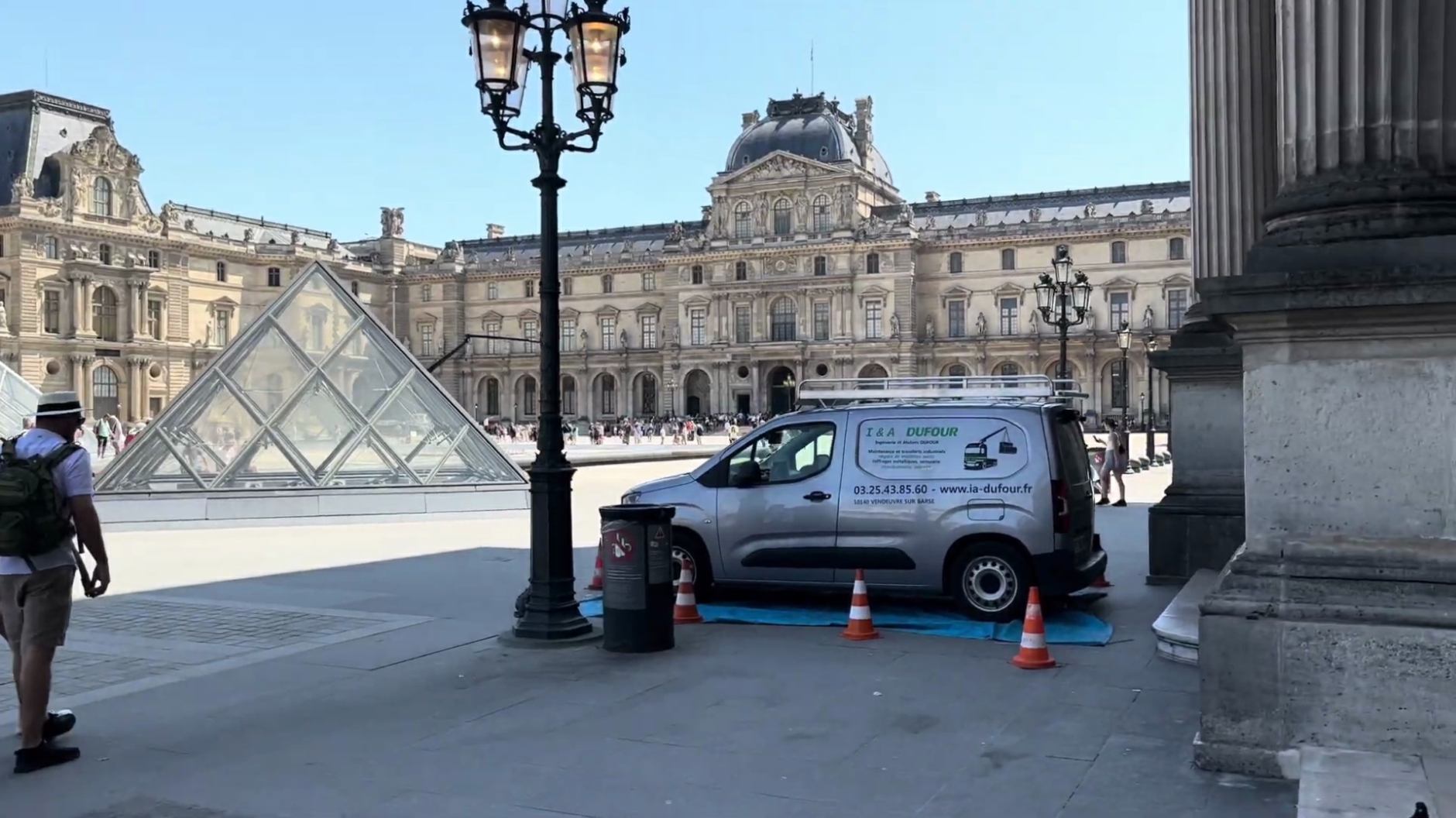}
    \includegraphics[width=0.108\textwidth]{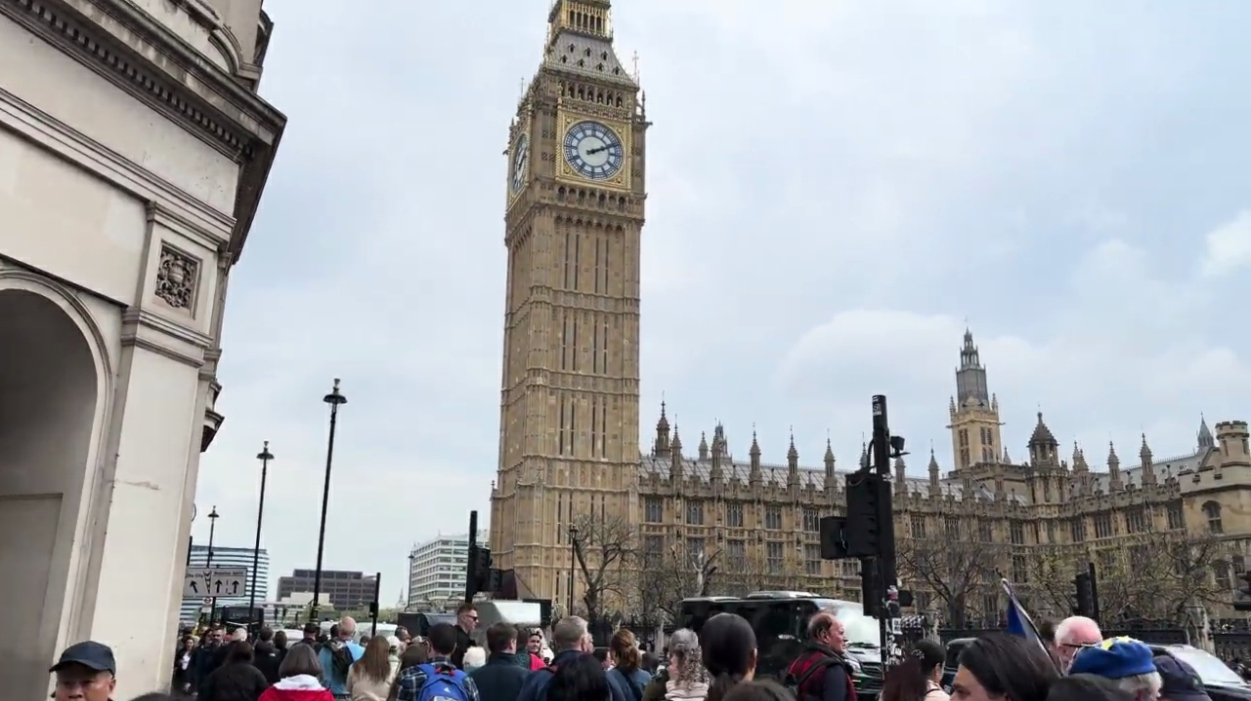}
    \\
    \caption{Some samples from each scene of our dataset. From left to right: \textbf{Perspolis, Eiffel Tower, Louvre Museum, and Big Ben} scenes.}
    \label{fig:dataset_overview}
\end{figure}
Please refer to Fig \ref{fig:dataset_overview} for some sample images from each of these scenes.

\section{Main object extraction}
\label{sec:main_obj_extraction}
\begin{figure*}[tbh] \centering
    \makebox[0.01\textwidth]{}
    \makebox[0.24\textwidth]{\scriptsize Garden Scene}
    \makebox[0.24\textwidth]{\scriptsize Statue Scene}
    \makebox[0.24\textwidth]{\scriptsize Truck Scene}
    \makebox[0.24\textwidth]{\scriptsize Eiffel Tower Scene}
    \\
    \raisebox{0.1\height}{\makebox[0.01\textwidth]{\rotatebox{90}{\makecell{\scriptsize $c_i \ge 0$}}}}
    \includegraphics[width=0.24\textwidth]{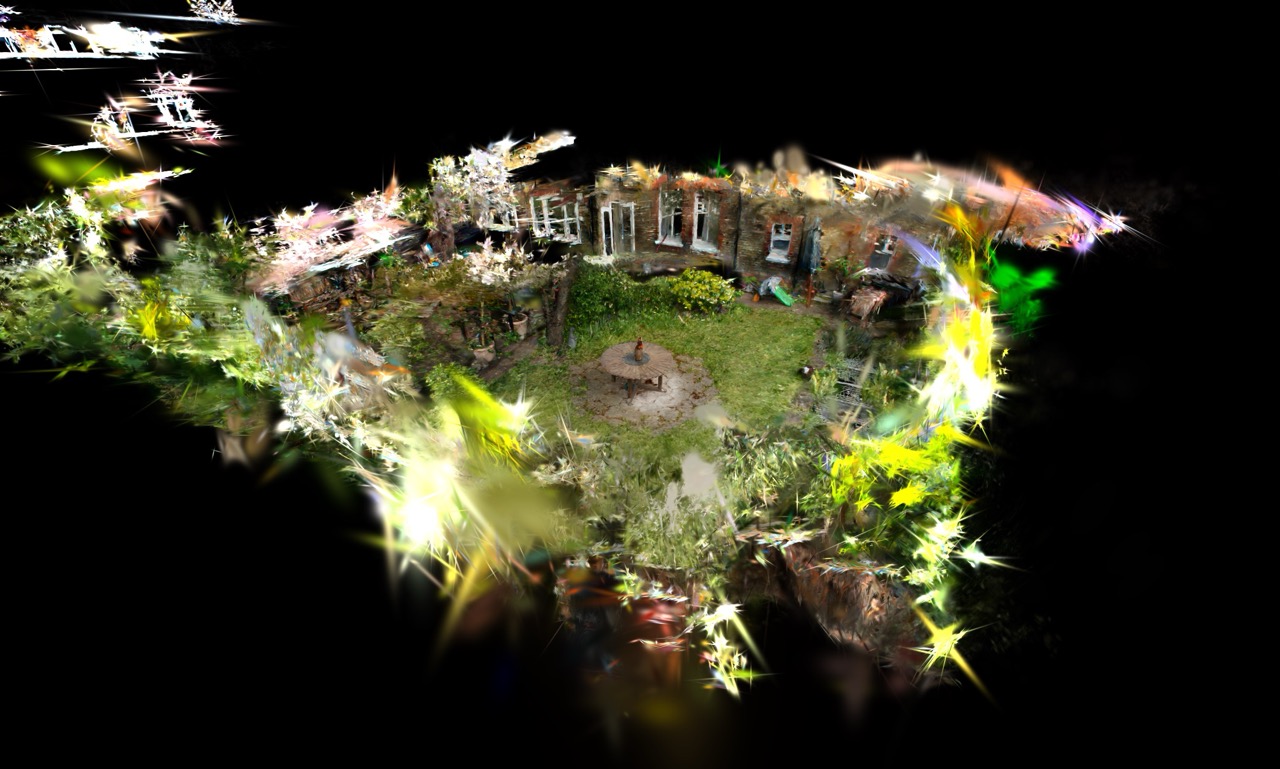}
    \includegraphics[width=0.24\textwidth]{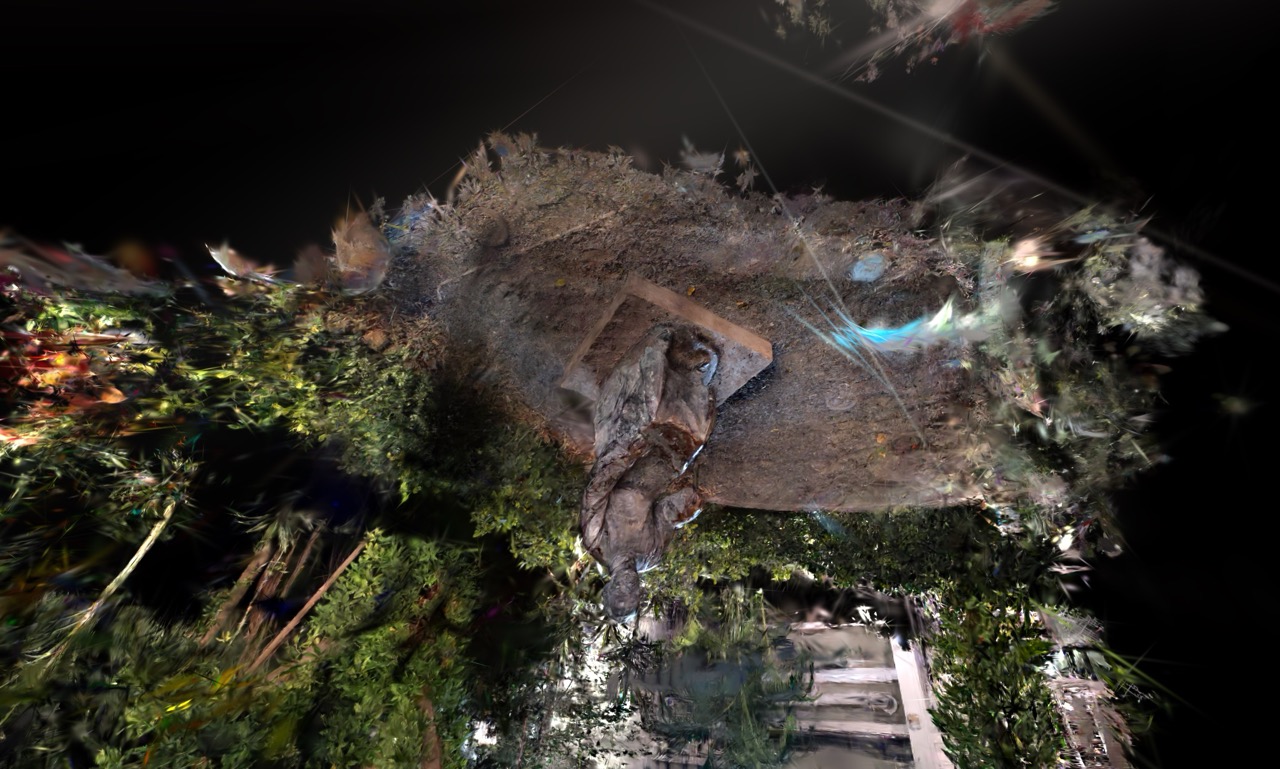}
    \includegraphics[width=0.24\textwidth]{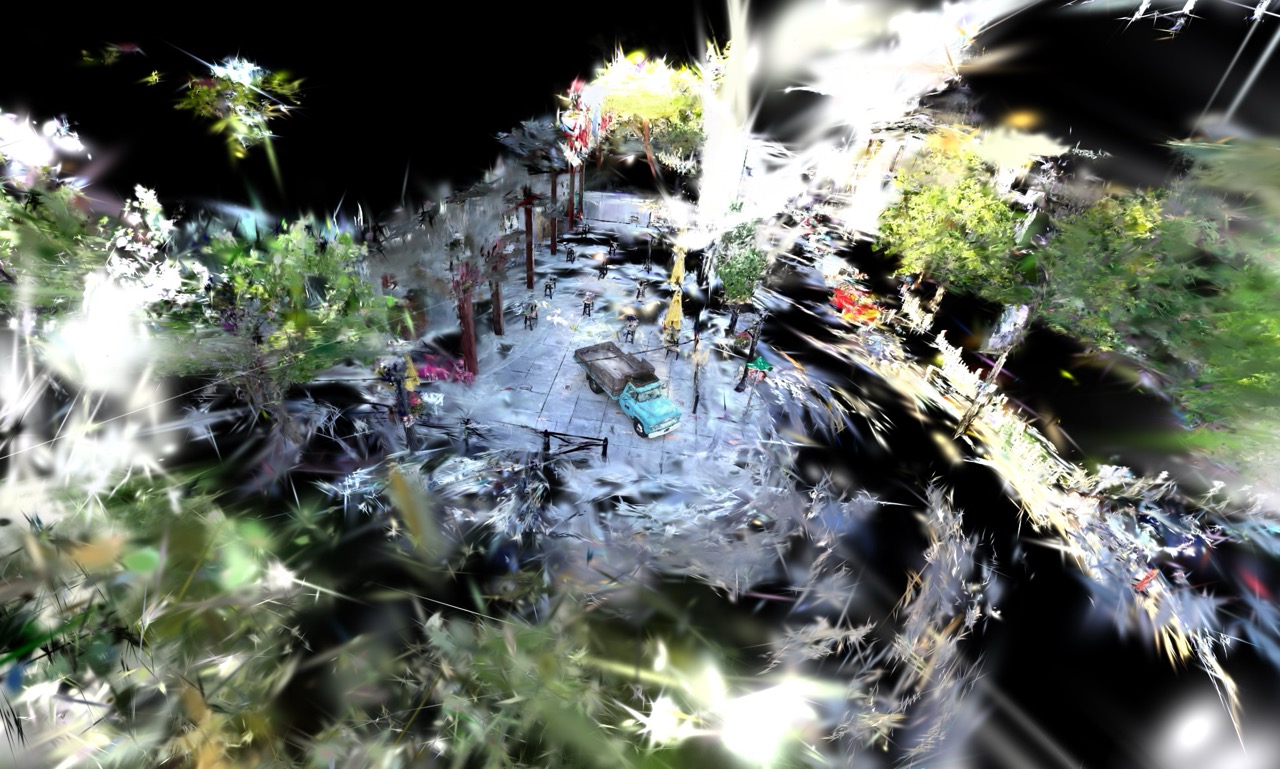}
    \includegraphics[width=0.24\textwidth]{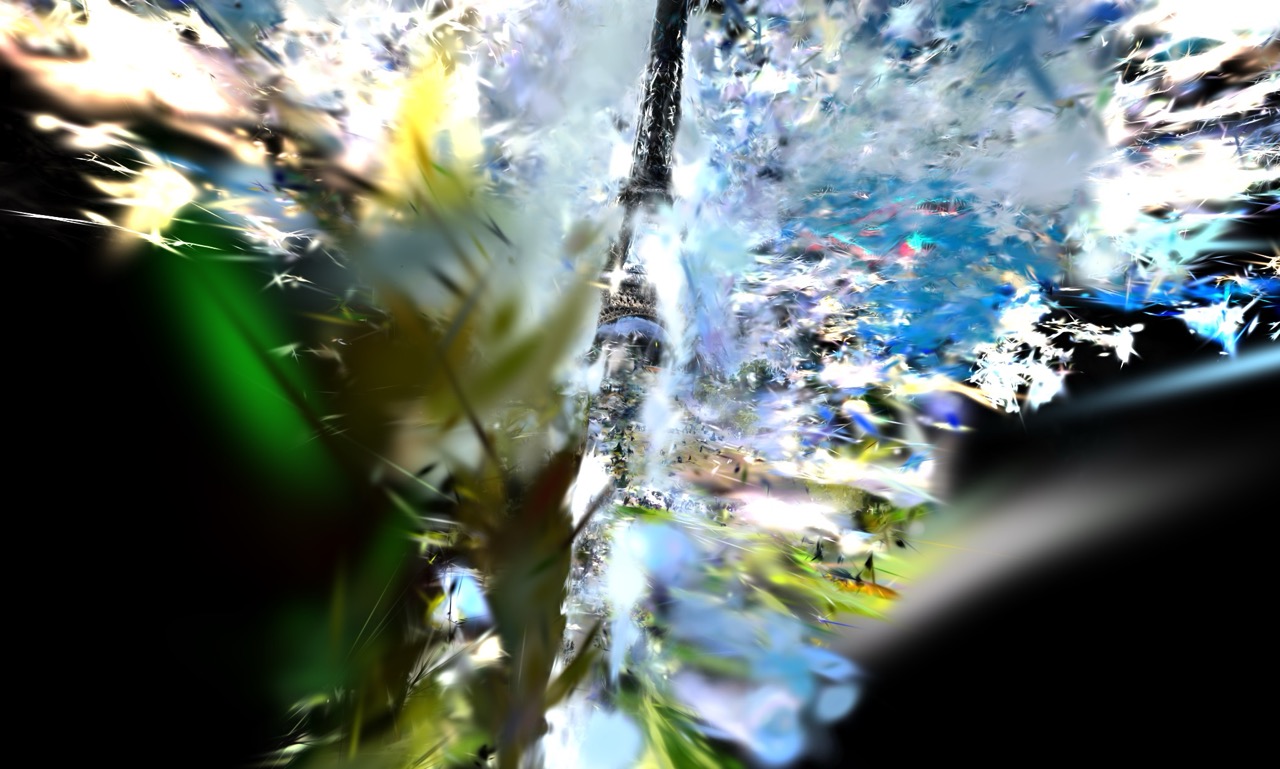}
    \\
    \raisebox{0.1\height}{\makebox[0.01\textwidth]{\rotatebox{90}{\makecell{\scriptsize $c_i \ge 0.5$}}}}
    \includegraphics[width=0.24\textwidth]{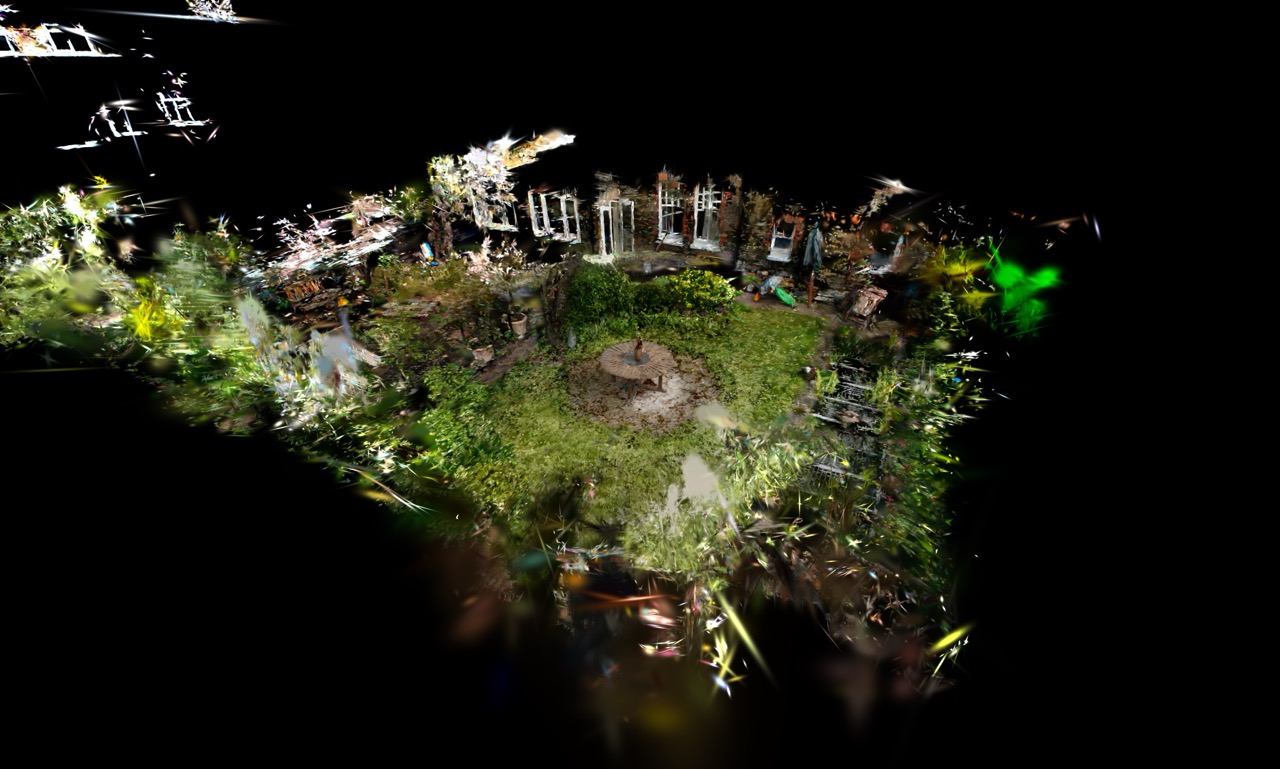}
    \includegraphics[width=0.24\textwidth]{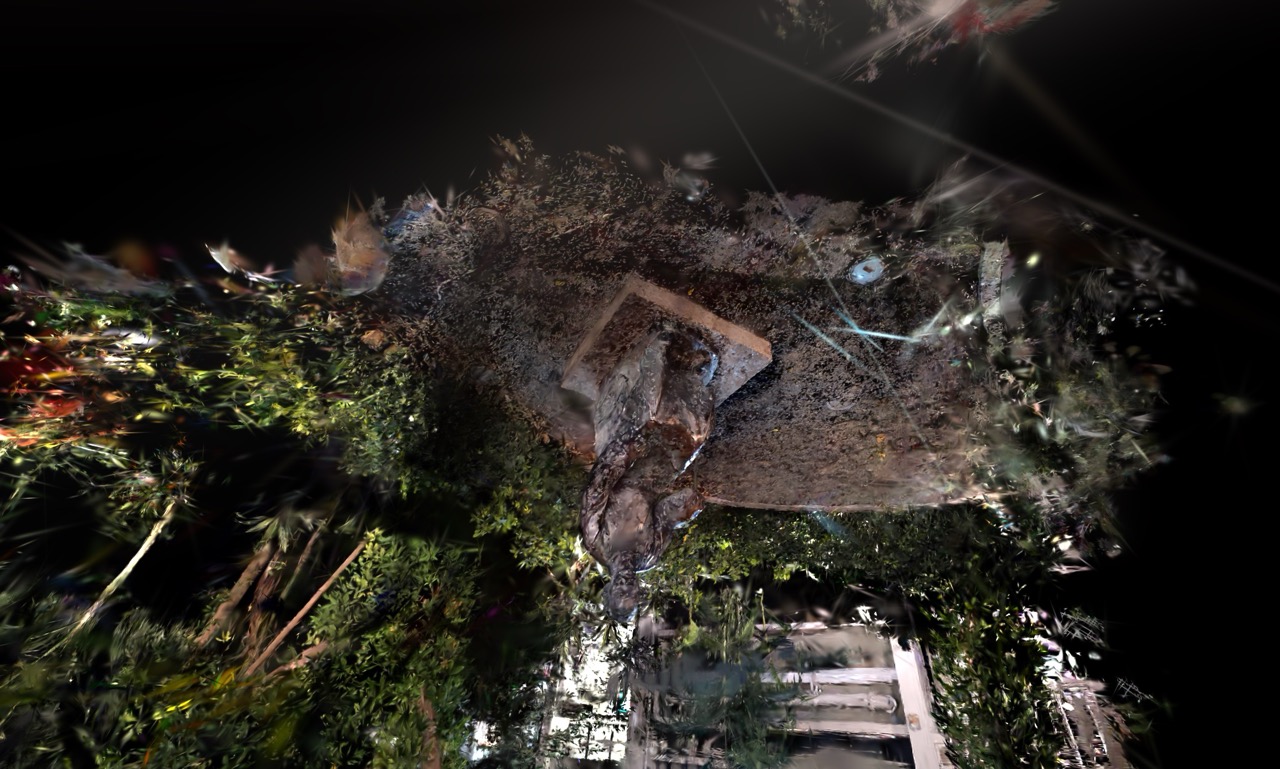}
    \includegraphics[width=0.24\textwidth]{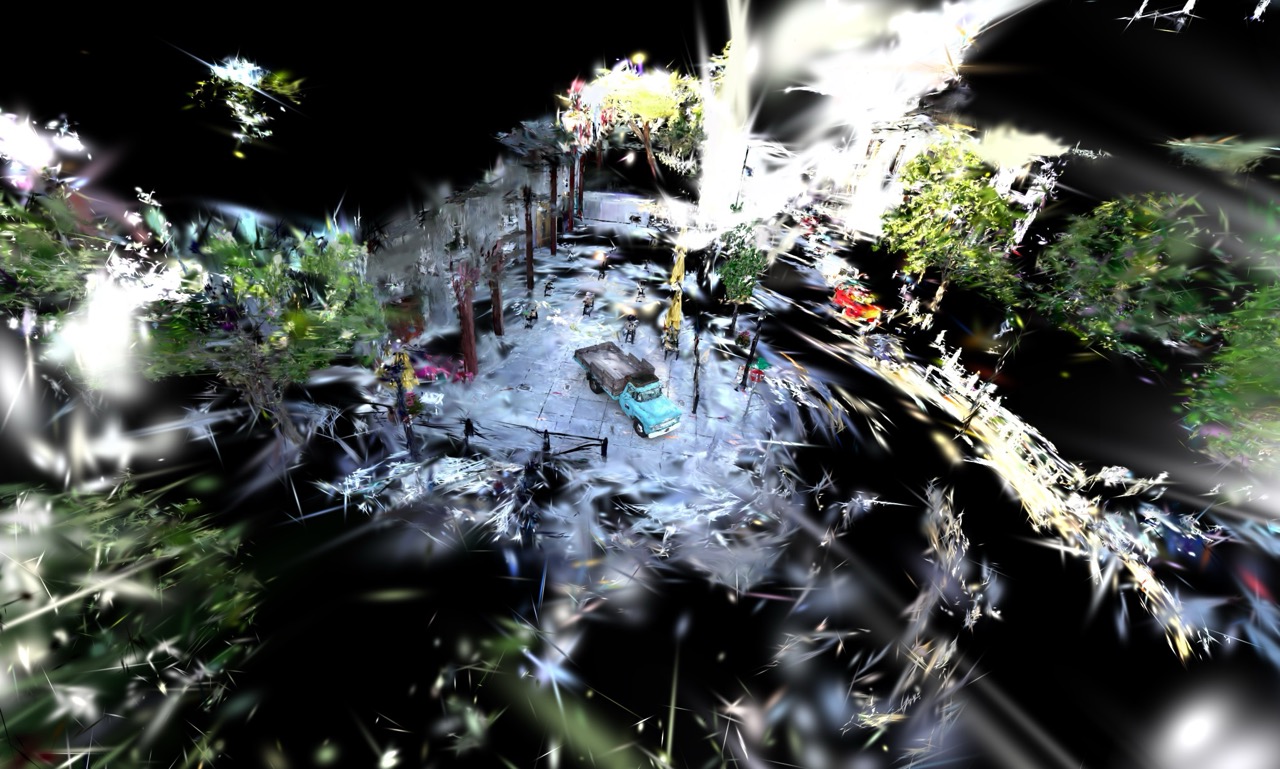}
    \includegraphics[width=0.24\textwidth]{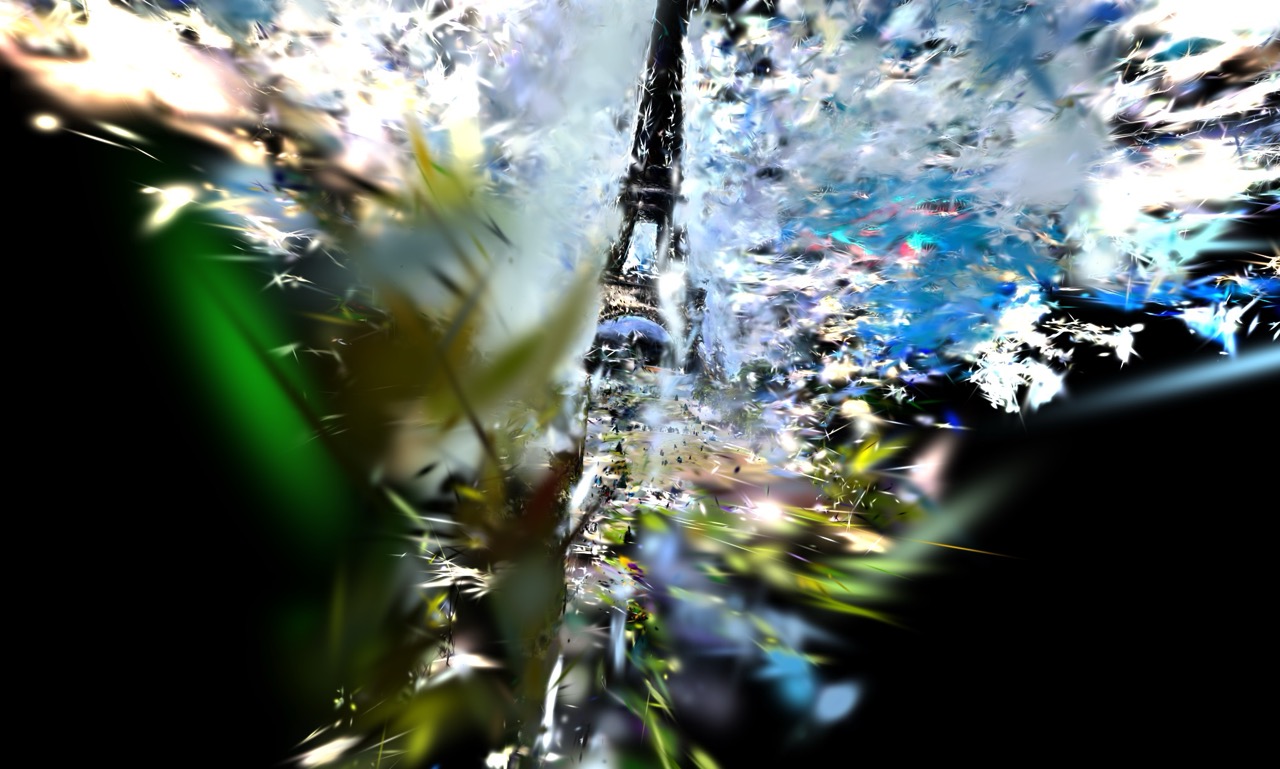}
    \\
    \raisebox{0.1\height}{\makebox[0.01\textwidth]{\rotatebox{90}{\makecell{\scriptsize $c_i \ge 0.75$}}}}
    \includegraphics[width=0.24\textwidth]{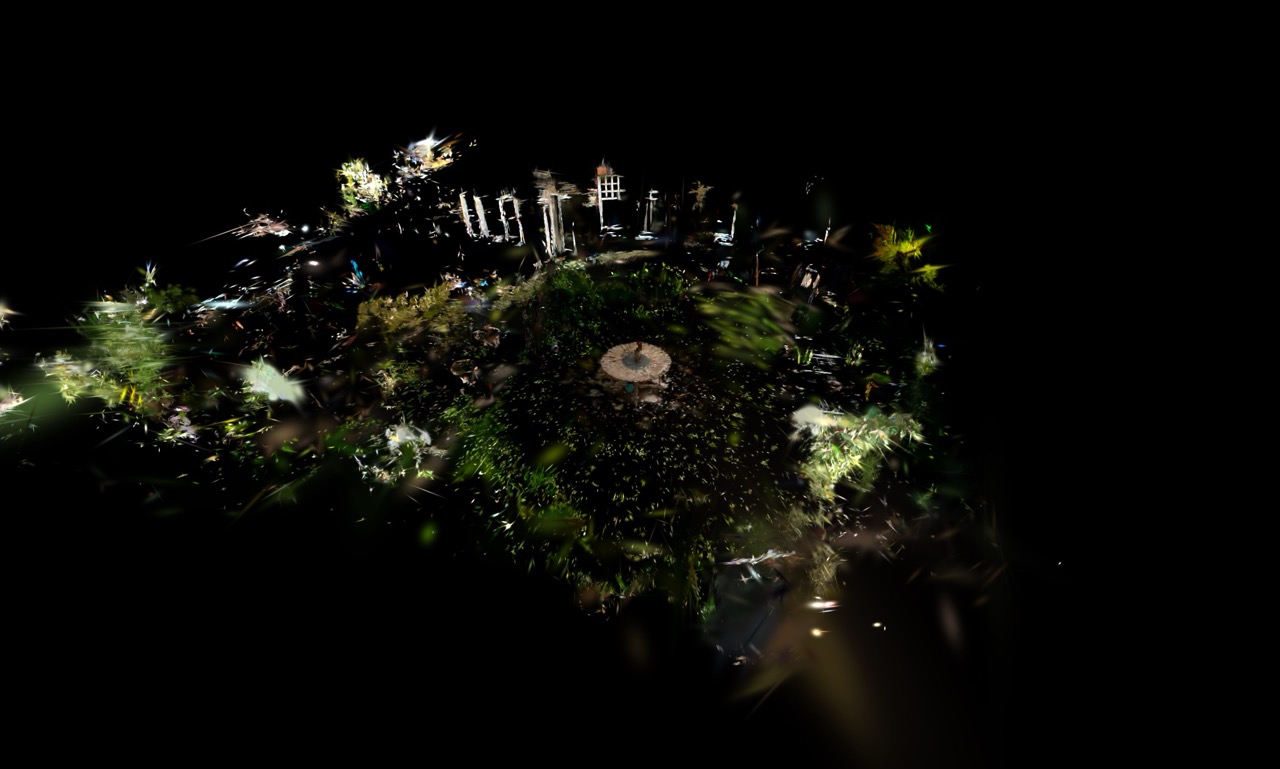}
    \includegraphics[width=0.24\textwidth]{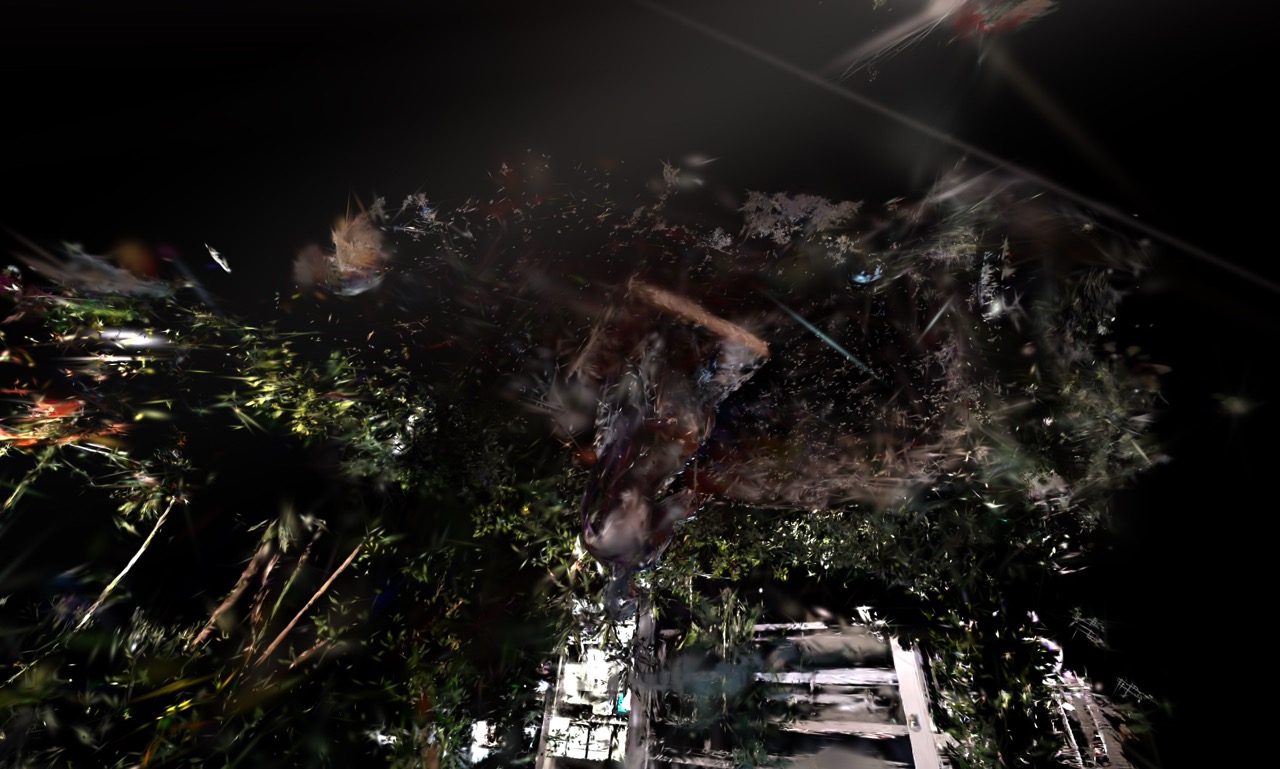}
    \includegraphics[width=0.24\textwidth]{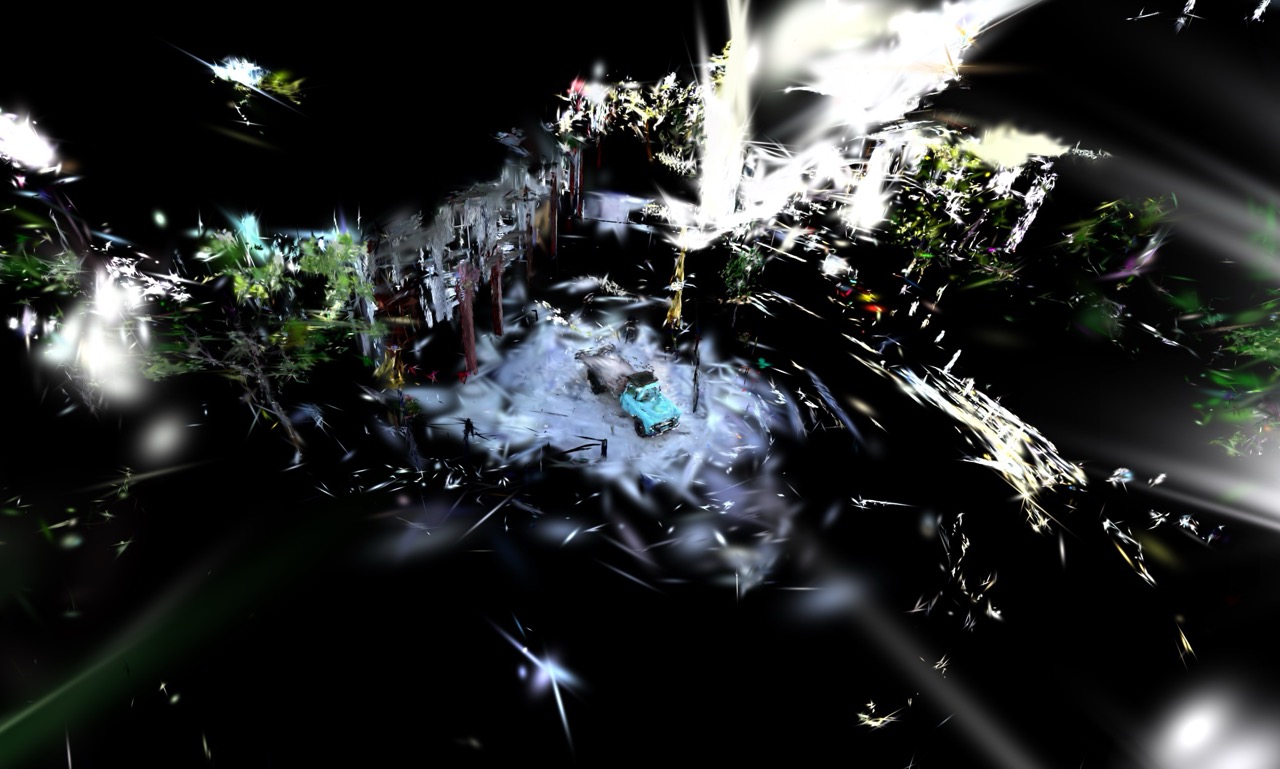}
    \includegraphics[width=0.24\textwidth]{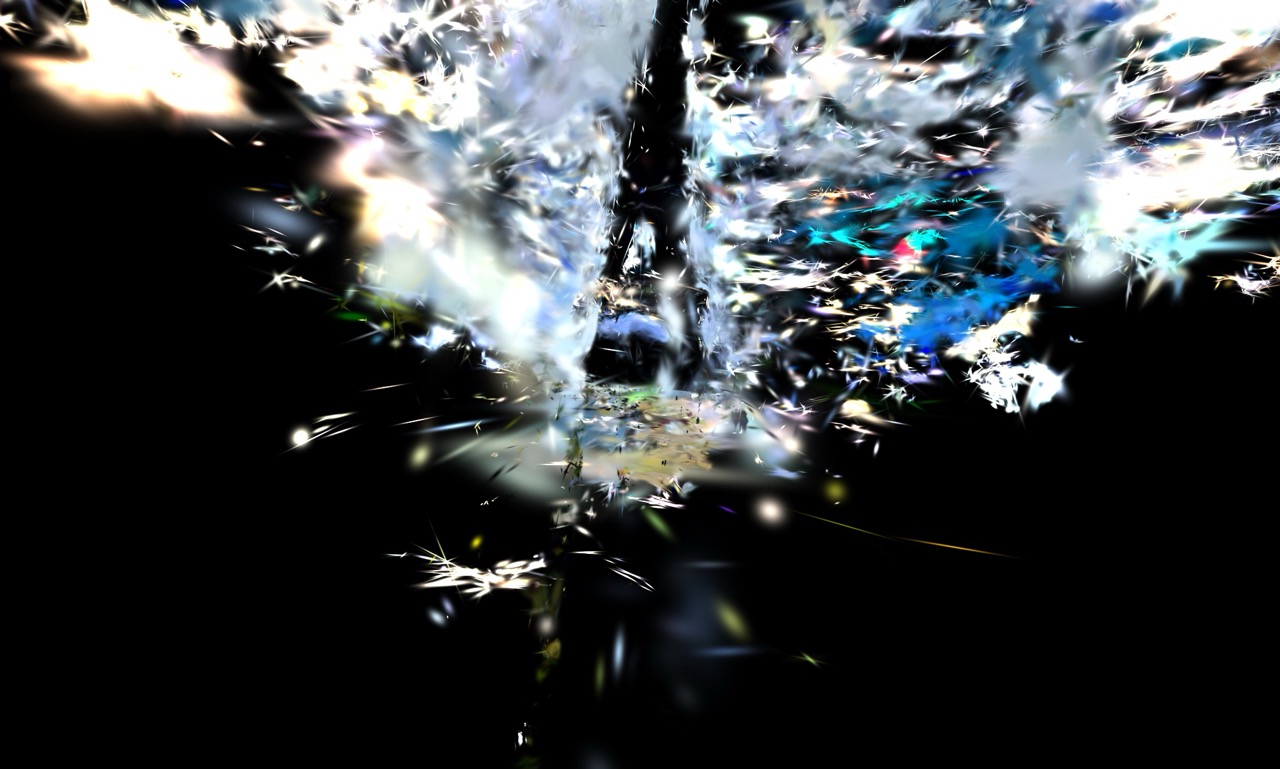}
    \\
    \caption{Visualizations among different scenes (horizontal axis) and different confidence thresholds (vertical axis). Note that these captures are from a perspective view (which are not in any of the test or train cameras) in our customized viewer. These scenes are intentionally shown from a high distance (like birds-eye view) to capture all floaters even outside our train/test camera frustums. You can see how many of these floaters are removed by increasing our confidence threshold.}
    \label{fig:floater_observations}
\end{figure*}

One of the usages of proposed method that can be studied further (in a separate work) is extracting the main object of the scene. Our results on various scenes show that when the confidence level is increased high enough, we get to a point that many of the surrounding objects get removed and the main object of the scene remains (For instance, refer to \ref{fig:garden_scene_qualititative} when threshold gets over $60\%$). By using a separate module on top of our confidence estimation, or even by enhancing our confidence estimation pipeline for the purpose of \textit{"Main object extraction"}, this task seems doable with high enough quality. As you can see in Fig \ref{fig:floater_observations}, by increasing confidence thresholds, the detail objects (like trees, environment objects, and background objects) are removed incrementally and the main object (alongside some other things, like confident reconstructed buildings) are remained.

\section{Floaters removal}
\label{sec:folaters_removal}
Another visionary for building on top of our pipeline is floater removal. This line of work is a famous problem in novel view synthesis and is being studied for a long time. Although the main aspect of our work i\textbf{s not} improving the scene quality, it can be used to remove the floaters (which can be interpreted as "\textit{low confidence points}" in the scene in comparison with others). For instance, refer to Fig \ref{fig:floater_observations} to see how different confidence thresholds result in removing different kind of floaters in the scene (inside our train/test camera views or outside them). 

\section{Effect of adding noise}
\label{sec:noise_addition_effect}
One of the methods for improving importance score (confidence score) prediction (which is a value between $0$ and $1$) is to add some multiplicative noise values. One of the most recent proposals in this area is LP-3DGS\cite{lp3dgs} paper which uses \textbf{Gumbel Noise} in a multiplicative way with its importance score and then applying a sigmoid function. A similar approach after calculating the expected value of confidence for each splat (which is $\frac{\alpha}{\alpha + \beta}$) is studied in Confident Splatting. Two cases have been experimented:
\begin{equation}
    c_i = Sigmoid( \frac{\hat{\alpha}}{\hat{\alpha} + \hat{\beta}} \times \frac{Gumbel\_noise}{\tau})
\end{equation}

Here, the Gumbel noise is added as a multiplicative noise for our estimated confidence and then applied Sigmoid to make the final prediction a smooth prediction between $[0, 1]$. The $\tau$ parameter is the temperature value for Gumbel noise, which is set to $2$ or $3$.

\begin{equation}
    c_i = Sigmoid( \frac{\hat{\alpha}}{\hat{\alpha} + \hat{\beta}} + \frac{Gumbel\_noise}{\tau})
\end{equation}
Here, the Gumbel noise is added as an additive noise for the estimated confidence and then Sigmoid is applied to make the final prediction a smooth prediction between $[0, 1]$.

While the \textit{additive version} results in better estimations than the \textit{multiplicative version}, their results are \textbf{not better} than the original predictions using raw $c_i = \frac{\hat{\alpha}}{\hat{\alpha} + \hat{\beta}}$. So, we chose to stick with the simpler, yet more powerful version of our pipeline that does not use the additive Gumbel noise.

\section{A more in-depth comparison}
\label{sec:in_depth_comparison}
In Table \ref{tab:comparison_complete} you can see a more comprehensive comparison of our model (added on top of original 3D gaussian splatting \cite{gaussian-splatting} and also Gaussian Splatting MCMC \cite{mcmc}) alongside the original 3DGS and also other pruning methods like RadSplat \cite{radsplat} and Mini-Splatting \cite{mini-splatting} with pruning ratio set to $50\%$. Please note the important difference between our method and these other methods, which is that our method is doing the pruning via thresholding in \textbf{test-time}, and thus, can adapt to any desired ratio that user wants based on the quality-compression tradeoff. But these two methods use a \textbf{training-time} pruning, so whenever user wants to change the pruning ratio, they would need to re-train the scene with new parameter.

\begin{table*}[tbh]
    \centering
    \resizebox{\linewidth}{!}{%
    \begin{tabular}{llccccc}
        \toprule
        \textbf{Scene} & \textbf{Method} & \textbf{PSNR ↑} & \textbf{SSIM ↑} & \textbf{LPIPS ↓} & \textbf{\#Gaussians ↓} & \textbf{SQR↓} \\
        \midrule
        \multirow{8}{*}{\textbf{Flowers}}
            & 3DGS        & $21.450$ & $0.596$ & $0.345$ & $3,590,000$ & $0.1433$ \\
            & Rad-Splat @ 50\%       & $21.367$ & $0.589$ & $0.361$ & $1,751,775$ & $0.0757$ \\
            & Mini-Splatting @ 50\%  & $21.382$ & $0.5871$ & $0.360$ & $3,384,239$ & $0.1366$ \\
            & Ours\texttt{@}Base3DGS @100\%    & $21.383$ & $0.5907$ & $0.360$ & $4,572,755$ & $0.1761$ \\
            & Ours\texttt{@}Base3DGS @$\sim$50\%     & $21.383$ & $0.5906$ & $0.360$ & $2,227,814$ & $0.0943$ \\
            & Ours\texttt{@}MCMC @100\%    & \textbf{21.592} & \textbf{0.6050} & \textbf{0.341} & $1,000,000$ & $0.0442$ \\
            & Ours\texttt{@}MCMC @$\sim$95\%    & $21.334$ & $0.5941$ & $0.347$ & $963,860$ & $0.0432$ \\
            & \textbf{Ours\texttt{@}MCMC @$\sim$90\%}    & $20.614$ & $0.5684$ & $0.360$ & \textbf{915,488} & \textbf{0.0425} \\
        \midrule
        
        \multirow{8}{*}{\textbf{Eiffel Tower}} 
            & 3DGS        & $22.877$ & $0.7969$ & $0.138$ & $600,000$ & $0.2077$ \\
            & Rad-Splat @50\%       & $22.524$ & $0.7919$ & $0.281$ & \textbf{395,873} & $0.1484$ \\
            & Mini-Splatting @50\%  & $22.189$ & $0.7884$ & $0.283$ & $803,662$ & $0.2658$ \\
            & Ours\texttt{@}Base3DGS @100\%    & $24.077$ & $0.843$ & $0.082$ & $1,496,522$ & $0.3833$ \\
            & \textbf{Ours\texttt{@}Base3DGS @$\sim$27\%}     & $23.552$ & $0.8329$ & $0.087$ & $408,564$ & \textbf{0.1478} \\
            & Ours\texttt{@}MCMC @100\%    & $24.653$ & $0.8585$ & $0.072$ & $1,000,000$ & $0.2885$ \\
            & Ours\texttt{@}MCMC @$\sim$95\%    & \textbf{24.653} & \textbf{0.8585} & \textbf{0.072} & $950,000$ & $0.2781$ \\
            & Ours\texttt{@}MCMC @$\sim$90\%    & $24.593$ & $0.8581$ & $0.072$ & $904,098$ & $0.2688$ \\
        \midrule
        \multirow{8}{*}{\textbf{Truck}} 
            & 3DGS        & $25.37$ & $0.8737$ & $0.099$ & $2,610,000$ & $0.0932$ \\
            & Rad-Splat @50\%       & $25.4187$ & $0.8818$ & $0.147$ & $1,291,794$ & $0.0483$ \\
            & Mini-Splatting @50\%  & $25.3142$ & $0.8801$ & $0.149$ & $2,555,598$ & $0.0916$ \\
            & Ours\texttt{@}Base3DGS @100\%    & $25.208$ & $0.8725$ & $0.103$ & $3,563,759$ & $0.1238$ \\
            & Ours\texttt{@}Base3DGS @$\sim$33\%     & $24.1741$ & $0.8564$ & $0.113$ & $1,142,117$ & $0.0451$ \\
            & Ours\texttt{@}MCMC @100\%   & \textbf{25.8155} & \textbf{0.8840} & \textbf{0.088} & $1,000,000$ & $0.0372$ \\
            & Ours\texttt{@}MCMC @$\sim$95\%    & $25.6068$ & $0.8801$ & $0.091$ & $950,877$ & $0.0358$ \\
            & \textbf{Ours\texttt{@}MCMC @$\sim$90\%}   & $24.7306$ & $0.8608$ & $0.107$ & \textbf{895,876} & \textbf{0.0349} \\
        \midrule
        \multirow{9}{*}{\textbf{Train}} 
            & 3DGS        & $21.5736$ & $0.807$ & $0.1635$ & $1,127,221$ & $0.0496$ \\
            & Rad-Splat @50\%       & $21.9755$ & $0.8133$ & $0.2101$ & $543,573$ & $0.0241$ \\
            & Mini-Splatting @50\%  & $21.6474$ & $0.8085$ & $0.2161$ & $1,043,839$ & $0.0460$ \\
            & Ours\texttt{@}Base3DGS @100\%    & $21.7970$ & $0.8017$ & $0.1764$ & $1,318,148$ & $0.0570$ \\
            & Ours\texttt{@}Base3DGS @$\sim$39\%     & $20.277$ & $0.7466$ & $0.216$ & $518,527$ & $0.0249$ \\
            & \textbf{Ours\texttt{@}Base3DGS @$\sim$34\%}     & $19.4559$ & $0.7136$ & $0.2413$ & \textbf{459,986} & \textbf{0.0230} \\
            & Ours\texttt{@}MCMC @100\%   & \textbf{22.6075} & \textbf{0.8285} & \textbf{0.1445} & $1,000,000$ & $0.0423$ \\
            & Ours\texttt{@}MCMC @$\sim$95\%    & $22.4773$ & $0.8253$ & $0.1447$ & $956,764$ & $0.0408$ \\
            & Ours\texttt{@}MCMC @$\sim$90\%   & $22.0062$ & $0.8086$ & $0.1563$ & $902,820$ & $0.0394$ \\
        \midrule
        \multirow{8}{*}{\textbf{Treehill}} 
            & 3DGS        & $22.4794$ & $0.6297$ & $0.3326$ & $3,742,946$ & $0.1427$ \\
            & Rad-Splat @50\%       & $22.4880$ & $0.6367$ & $0.3505$ & $1,745,215$ & $0.0720$ \\
            & Mini-Splatting @50\%  & $22.4921$ & $0.6362$ & $0.3490$ & $3,301,114$ & $0.1279$ \\
            & Ours\texttt{@}Base3DGS @100\%    & $22.5340$ & $0.6308$ & $0.3421$ & $4,399,502$ & $0.1633$ \\
            & Ours\texttt{@}Base3DGS @$\sim$47\%     & $21.5555$ & $0.5997$ & $0.3723$ & $2,092,794$ & $0.0884$ \\
            & Ours\texttt{@}MCMC @100\%   & $22.9600$ & $0.6456$ & $0.3520$ & $1,000,000$ & $0.0417$ \\
            & Ours\texttt{@}MCMC @$\sim$95\%    & $22.5186$ & $0.6348$ & $0.3598$ & $955,394$ & $0.0406$ \\
            & \textbf{Ours\texttt{@}MCMC @$\sim$90\%}   & $21.3151$ & $0.6002$ & $0.3870$ & $884,206$ & \textbf{0.0398} \\
        \bottomrule
    \end{tabular}
    }
    \vspace{0.5em}
    \caption{A complete view of our comparisons among different methods and pruning ratios. @$X\%$ means that specific method, is keeping $X\%$ of the splats initially in its scene.}
    \label{tab:comparison_complete}
\end{table*}

\end{document}